\documentclass[pdflatex,sn-nature,iicol]{sn-jnl}

\usepackage{graphicx}
\usepackage{amssymb,amsmath,amsfonts}
\usepackage{booktabs}
\usepackage{multirow}
\usepackage{etoolbox}
\usepackage{flafter}
\usepackage{capt-of}
\usepackage{stfloats}
\usepackage{placeins}
\usepackage{needspace}
\usepackage{balance}

\makeatletter
\providecommand\@makespecialcolbox{\@make@specialcolbox}
\makeatother

\begin{document}

\title{NEP-EM26: A Neuroevolution Potential for Accurate and Efficient Detonation Velocity Prediction in CHNO Energetic Materials} 
\author[1]{\fnm{Zhenkun} \sur{Qiu}}
\author[1]{\fnm{Junwen} \sur{Duan}}
\author[1]{\fnm{Zifeng} \sur{Liu}}
\author*[2]{\fnm{Zheyong} \sur{Fan}}\email{brucenju@gmail.com}
\author*[1,3,4]{\fnm{Xuecheng} \sur{Shao}}\email{shaoxc@jlu.edu.cn}
\author*[1,3,4]{\fnm{Wenhui} \sur{Mi}}\email{mwh@jlu.edu.cn}

\affil[1]{
  \orgdiv{Key Laboratory of Material Simulation Methods \& Software of Ministry of Education, College of Physics},
  \orgname{Jilin University},
  \orgaddress{
    \city{Changchun},
    \postcode{130012},
    \country{China}}}

\affil[2]{
  \orgdiv{College of Physical Science and Technology},
  \orgname{Bohai University},
  \orgaddress{
    \city{Jinzhou},
    \postcode{121013},
    \country{China}}}

\affil[3]{
  \orgdiv{State Key Laboratory of High Pressure and Superhard Materials, College of Physics},
  \orgname{Jilin University},
  \orgaddress{
    \city{Changchun},
    \postcode{130012},
    \country{China}}}

\affil[4]{
  \orgdiv{International Center of Future Science},
  \orgname{Jilin University},
  \orgaddress{
    \city{Changchun},
    \postcode{130012},
    \country{China}}}

\abstract{
Accurate and efficient prediction of detonation velocities across diverse energetic materials requires an interatomic potential capable of describing thermodynamic states ranging from unreacted materials under ambient conditions to detonation products at high temperatures and pressures. Here, we develop NEP-EM26, a neuroevolution potential for calculating the detonation velocities of CHNO energetic materials, trained on density functional theory (DFT) data spanning diverse crystal structures and reactive configurations. Validation against DFT demonstrates that NEP-EM26 reproduces the static energy--volume curves of eight tested CHNO explosives and captures the local pair structures and pressure--volume behavior of reacting CL-20 and RDX. Across these eight explosives, including four without dedicated reactive molecular dynamics (MD) training configurations, NEP-EM26 predicts detonation velocities with a mean absolute percentage error of 2.4\% relative to experiment, approximately one-third that of ReaxFF (7.4\%). Moreover, NEP-EM26 maintains an absolute percentage error below 5\% for every material. Computational benchmarks further show that NEP-EM26 in GPUMD runs more than ten times faster per MD step than ReaxFF in LAMMPS on the same GPU. NEP-EM26 thus bridges DFT-based atomistic modeling and accurate, efficient, and transferable detonation-velocity prediction for CHNO energetic materials.
}

\keywords{machine-learning interatomic potential, neuroevolution potential,
CHNO energetic materials, reactive molecular dynamics, Hugoniot of detonation products,
detonation velocity}

\maketitle

%%=============================================================
\section{Introduction}
\label{sec:intro}

Energetic materials are widely used in mining, detonation propulsion, and defense owing to their high detonation performance~\cite{kamlet1968,handley2018,hamilton2021_llm105,perut2007}. Detonation velocity is a key metric of this performance, making its accurate measurement and prediction central to the design and evaluation of energetic materials. However, experimental measurement of detonation velocity requires sufficient sample quantities and carefully controlled tests, rendering it both time-consuming and costly~\cite{elton2018_ml_energetic}. Theoretical prediction is therefore indispensable for evaluating detonation velocity, particularly during the early-stage development of novel energetic materials when only limited material is available for experimental testing. Theoretical approaches for predicting detonation velocity generally fall into three categories. Empirical relations, such as the Kamlet--Jacobs equations, yield rapid estimates from bulk properties~\cite{kamlet1968}. Thermochemical codes, such as CHEETAH and EXPLO5, offer a more detailed description of detonation products and are widely used for conventional CHNO energetic materials~\cite{cheetah1993,explo5_2001}, although their predictive reliability for novel compounds requires system-specific validation. Atomistic simulations based on molecular dynamics (MD) provide a complementary route by explicitly resolving interatomic interactions and chemical reactions~\cite{jadrich2021_eos,deepmd_cj2023}. This approach establishes a general framework for predicting detonation velocities across diverse classes of energetic materials, with both accuracy and computational efficiency governed primarily by the underlying interatomic potential.

Within this framework, MD simulations can either track shock propagation directly or sample shocked states subject to the Rankine--Hugoniot conservation relations, as in the multiscale shock technique and Hugoniostat methods~\cite{reed2003msst,maillet2000nphug,wen2022_shock_tutorial}. In pressure-controlled Hugoniostat simulations, the system evolves under the Hugoniot energy constraint to sample thermodynamic states at prescribed pressures~\cite{ravelo2004nphug}. Repeating these simulations across a range of pressures yields the detonation-product Hugoniot, and the Chapman--Jouguet (CJ) state is determined by the tangency point of the Rayleigh line passing through the initial state~\cite{jadrich2021_eos}. Obtaining reliable detonation-product Hugoniot states requires chemical and thermodynamic equilibration over a broad range of conditions, which makes first-principles simulations computationally demanding. DFT-MD is typically limited to picosecond-scale trajectories, whereas reaction-zone timescales in common explosives can extend to nanoseconds~\cite{jadrich2021_eos,pham2020_hn3}. To mitigate this bottleneck, the Decoupled Hugoniostat (D-Hug) framework initiates simulations from product-like configurations and reduces the required equilibration time by more than two orders of magnitude (Z. Liu et al., manuscript in preparation). Even with this acceleration, systematic evaluations across multiple materials still require reactive-MD trajectories at several pressures for each system. Moreover, errors in the energetics of the reaction products can propagate into the predicted CJ velocity, pressure, and temperature, as demonstrated for pentaerythritol tetranitrate (PETN)~\cite{jadrich2021_eos}. Accordingly, accurate and efficient CJ detonation-velocity prediction requires a computationally affordable interatomic potential that captures the energetics of both unreacted materials under ambient conditions and reactive states across the wide temperature and pressure ranges of detonation.

Reactive force fields, including empirical models such as ReaxFF~\cite{vanduin2001,strachan2003rdx} and neural-network reactive force fields~\cite{hamilton2023_nnrf}, substantially reduce computational cost by describing bond breaking and formation and have been widely applied to energetic materials. However, their accuracy depends strongly on the chosen functional form and parameterization, which can limit reliability outside the fitted chemical and thermodynamic regimes~\cite{hamilton2023_nnrf,islam2019_nm}. Machine-learning interatomic potentials (MLIPs) offer a promising alternative by learning energies and forces directly from DFT reference data at a fraction of the first-principles cost, and recent MLIPs have demonstrated rapid progress in this field. For example, material-specific Deep Potential models have been used to predict the CJ properties of CL-20 and NTO~\cite{deepmd_cj2023,nto_dnnp_2024}. Expanding beyond single compounds, the multi-material EMFF-2025 potential addresses the structural, mechanical, and decomposition properties of 20 energetic materials~\cite{emff2025}, while general-purpose CHNO potentials have been developed to capture diverse reaction pathways, energies, and barriers~\cite{li2025chon_reactive}. Although these studies establish the viability of MLIPs for reactive energetic materials, developing a single transferable potential across the full thermodynamic range of detonation remains a central challenge. Specifically, such a potential must accurately describe both unreacted materials under ambient conditions and extreme detonation states reaching several thousand kelvin and pressures exceeding 100~GPa~\cite{pham2026_shocked_mlff}, while achieving high computational efficiency and keeping detonation-velocity prediction errors comparable to experimental uncertainties~\cite{suceska2025_exp6}.

The neuroevolution potential (NEP) framework is particularly well suited for this task because its compact neural-network architecture and GPU-optimized implementation in GPUMD enable highly efficient atomistic simulations~\cite{fan2021nep,fan2022gpumd}. Previous NEP studies have investigated shock-induced hotspot formation, reaction kinetics, and decomposition in CL-20 and related crystals~\cite{hu2026cl20_hotspot,hu2025cl20_hostguest}. Extending NEP from individual material families to systematic, experimentally benchmarked detonation-velocity predictions across diverse CHNO explosives requires a unified potential that maintains accuracy over distinct physical and chemical environments without sacrificing computational speed.

Here, we develop NEP-EM26, a single DFT-trained NEP constructed from diverse crystal structures and high-temperature, high-pressure reactive configurations. We validate NEP-EM26 against held-out DFT data, static energy--volume curves of eight crystals, and local structural and pressure--volume properties at selected reactive states of CL-20 and RDX. Coupled with the D-Hug framework, NEP-EM26 predicts detonation velocities for eight CHNO explosives, including four without dedicated reactive-MD training configurations. Relative to experiment, NEP-EM26 achieves a mean absolute percentage error (MAPE) of 2.4\%, compared with 7.4\% for the Liu--Goddard ReaxFF parameterization~\cite{liu2011_reaxfflg} (hereafter ReaxFF), while running more than an order of magnitude faster per MD step than ReaxFF on the same GPU. Overall, these results demonstrate that a unified DFT-trained interatomic potential can bridge atomistic modeling and experimentally measurable detonation properties at a substantially reduced computational cost.

%%=============================================================
\section{Results}
\label{sec:results}
%%=============================================================

Our assessment of NEP-EM26 proceeds from configuration-level errors and static crystal responses
to reactive-state structure, pressure--volume behavior, and detonation
velocities, linking the DFT comparisons to detonation-velocity prediction.

\subsection{Energy, Force, and Stress Accuracy}
\label{sec:accuracy}

\begin{figure*}[!t]
\centering
\includegraphics[width=\textwidth]{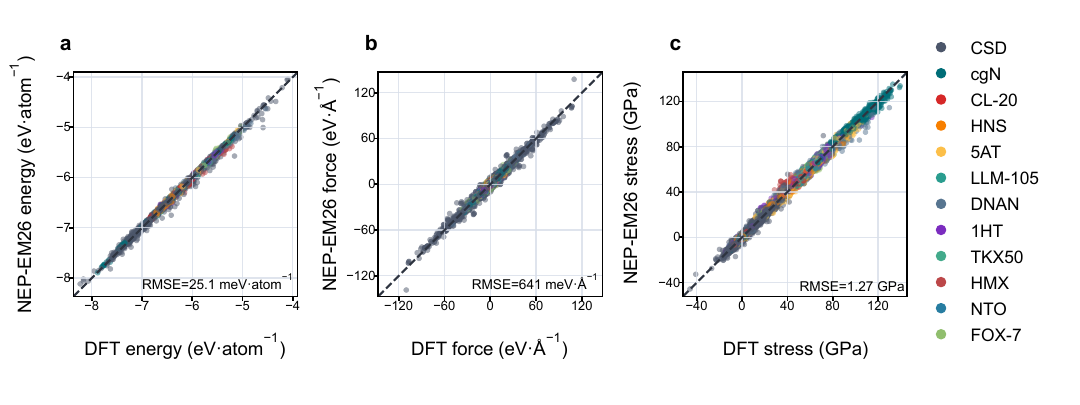}
\caption{NEP-EM26 predictions compared with DFT reference values for 8{,}151 held-out configurations.
\textbf{a}, Energy per atom; \textbf{b}, atomic force components; \textbf{c}, stress components (compression positive).
Dashed lines indicate exact agreement.
Colors distinguish the CSD structures and the eleven reactive-MD sources, whose abbreviations are defined in Table~\ref{tab:source_errors}.}
\label{fig:parity}
\end{figure*}

On the configuration-wise randomly held-out test set, NEP-EM26 achieves pooled root-mean-square errors (RMSEs) of
25.1\,meV$\cdot$atom$^{-1}$, 641\,meV$\cdot$\AA$^{-1}$, and 1.27\,GPa for
energy per atom, force components, and stress components, respectively
(Fig.~\ref{fig:parity}). The held-out test set contains 8{,}151 configurations:
5{,}503 CSD structures and 2{,}648 configurations from eleven reactive-MD
systems at high temperatures and pressures.

\begin{table*}[!t]
\centering
\begin{threeparttable}
\caption{Source-resolved root-mean-square errors on the held-out test set.
The reactive-MD sources are cgN, cubic gauche nitrogen;
CL-20, hexanitrohexaazaisowurtzitane;
HNS, 2,2$'$,4,4$'$,6,6$'$-hexanitrostilbene;
5AT, 5-aminotetrazole;
LLM-105, 2,6-diamino-3,5-dinitropyrazine-1-oxide;
DNAN, 2,4-dinitroanisole;
1HT, 1$H$-tetrazole;
TKX50, dihydroxylammonium 5,5$'$-bistetrazole-1,1$'$-diolate;
HMX, octahydro-1,3,5,7-tetranitro-1,3,5,7-tetrazocine;
NTO, 3-nitro-1,2,4-triazol-5-one; and
FOX-7, 1,1-diamino-2,2-dinitroethene.}
\label{tab:source_errors}
\renewcommand{\arraystretch}{1.06}
\begin{tabular*}{\textwidth}{@{\extracolsep{\fill}}ccccc@{}}
\toprule
\multirow{2}{*}{Source} &
\multirow{2}{*}{\shortstack{No. of test\\configurations}} &
\multicolumn{3}{c}{Held-out RMSE} \\
\cmidrule(lr){3-5}
& & \shortstack{Energy per atom\\(meV$\cdot$atom$^{-1}$)} &
\shortstack{Force component\\(meV$\cdot$\AA$^{-1}$)} &
\shortstack{Stress component\\(GPa)} \\
\midrule
CSD    & 5{,}503 & 25.0 & 342 & 0.83 \\
\addlinespace[2pt]
cgN    & 348     & 20.7 & 948 & 1.87 \\
CL-20   & 339     & 21.2 & 776 & 1.32 \\
HNS    & 295     & 22.3 & 903 & 1.84 \\
5AT    & 295     & 25.6 & 856 & 1.91 \\
LLM-105 & 306     & 20.9 & 808 & 1.68 \\
DNAN   & 245     & 33.2 & 989 & 2.76 \\
1HT    & 211     & 24.1 & 925 & 2.05 \\
TKX50  & 211     & 21.1 & 713 & 1.43 \\
HMX    & 162     & 42.5 & 799 & 2.24 \\
NTO    & 136     & 19.7 & 779 & 1.46 \\
FOX-7   & 100     & 31.7 & 886 & 2.03 \\
\bottomrule
\end{tabular*}
\end{threeparttable}
\end{table*}

The source-resolved errors reveal different patterns for energy, force, and
stress (Table~\ref{tab:source_errors}). The CSD subset has force and stress
RMSEs of 342\,meV$\cdot$\AA$^{-1}$ and 0.83\,GPa, respectively, both lower
than those of any reactive-MD source, but its energy RMSE
(25.0\,meV$\cdot$atom$^{-1}$) is higher than that of seven of the eleven
reactive-MD sources. Among the reactive-MD sources, HMX has the largest energy RMSE
(42.5\,meV$\cdot$atom$^{-1}$), whereas DNAN has the largest force and stress
RMSEs (989\,meV$\cdot$\AA$^{-1}$ and 2.76\,GPa). The pooled values thus
average over source-level errors that differ appreciably, particularly for force
and stress. Supplementary Fig.~S1 further resolves the reactive-MD errors
by temperature and pressure.

\subsection{Static Energy--Volume Comparisons}
\label{sec:eos}

\begin{figure*}[!htp]
\centering
\includegraphics[width=\textwidth]{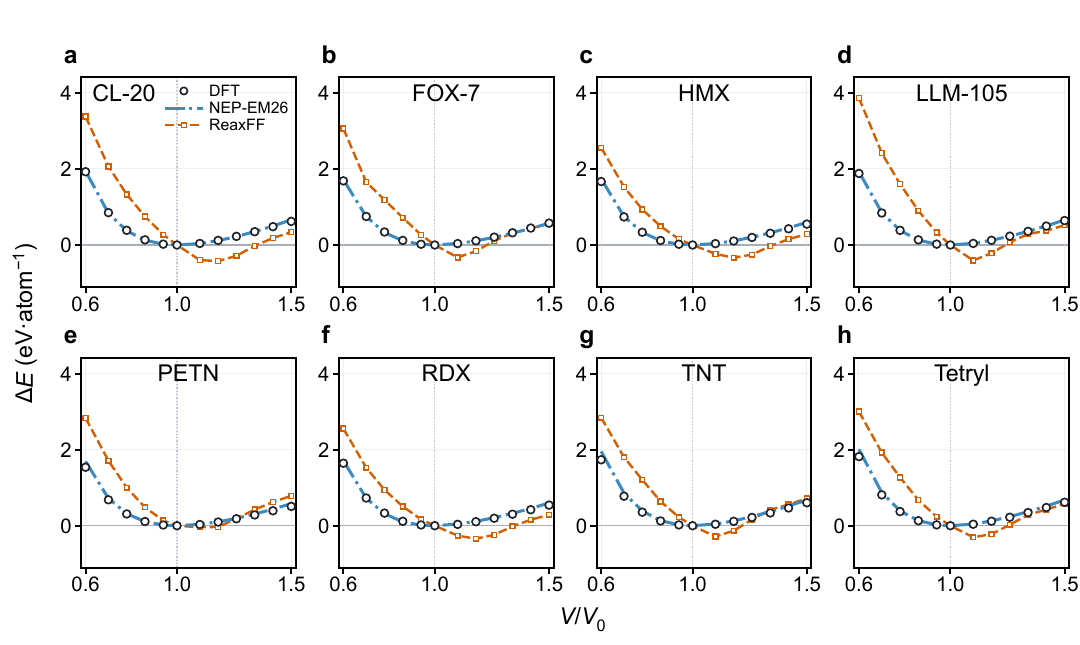}
\caption{Static energy--volume curves from DFT, NEP-EM26, and ReaxFF.
\textbf{a--d}, Reactive-MD source materials: CL-20, FOX-7, HMX, and LLM-105, respectively.
\textbf{e--h}, Materials without dedicated reactive-MD training configurations: PETN, RDX, TNT, and Tetryl, respectively.
Relative energy per atom $\Delta E$ is plotted against $V/V_0$, where $V_0$ is the DFT-optimized crystal volume.
Each curve is shifted to zero energy at $V/V_0=1$, indicated by the vertical reference line.
}
\label{fig:static_eos}
\end{figure*}

Static energy--volume curves probe the potential's response to systematic
changes in crystal volume. We compare these curves for the same eight crystals reported in the
detonation-velocity benchmark (Fig.~\ref{fig:static_eos}). Four of them
(CL-20, FOX-7, HMX, and LLM-105) have dedicated reactive-MD training
configurations, whereas the other four (PETN, RDX, TNT, and Tetryl) do not. The other seven
reactive-MD sources in Table~\ref{tab:source_errors} extend the training
environments beyond these benchmark materials. Each crystal is scaled isotropically over $V/V_0=0.60$--1.50, where $V_0$ is the DFT-optimized volume, and all three methods are evaluated on identical configurations (Methods).

NEP-EM26 reproduces both the steep rise of the DFT energy under compression and its
gentler rise under expansion for all eight crystals. Over all sampled volumes,
its $\Delta E$ RMSE relative to DFT is 42.1\,meV$\cdot$atom$^{-1}$, compared with
608.2\,meV$\cdot$atom$^{-1}$ for ReaxFF, and it is lower than the ReaxFF value for every one of the eight materials. Supplementary Fig.~S2 shows the corresponding comparison for an earlier ReaxFF
parameterization.

By material group, the RMSE is 17.4\,meV$\cdot$atom$^{-1}$ for the four
reactive-MD source materials and 57.0\,meV$\cdot$atom$^{-1}$ for the other
four (Supplementary Table~S2).
Because cell vectors and atomic coordinates are scaled together, compression
also shortens covalent bonds, so the most compressed points probe
intramolecular repulsion as well as intermolecular packing. The largest
deviations occur at $V/V_0=0.60$, for TNT, Tetryl, and PETN; these points
account for 71\% of the NEP-EM26 squared error. Over $V/V_0=0.78$--1.26,
excluding the reference point, the NEP-EM26 RMSE is 9.4\,meV$\cdot$atom$^{-1}$,
compared with 483.4\,meV$\cdot$atom$^{-1}$ for ReaxFF.
NEP-EM26 thus retains DFT-like
curve shapes beyond its reactive-MD source materials, with larger
residuals for some crystals at the strongest compression.

\subsection{Local Structure and Dynamics in CL-20 and RDX}
\label{sec:rdf}

Reactive trajectories extend the assessment from static crystal responses
to the configurations reached during high-pressure reactions.
We compare Hugoniostat
trajectories from DFT, NEP-EM26, and ReaxFF for 144-atom CL-20 and 168-atom RDX systems at approximately
46 and 33\,GPa, respectively, with all three methods using the same initial
configuration.

\begin{figure*}[!htp]
\centering
\includegraphics[width=\textwidth]{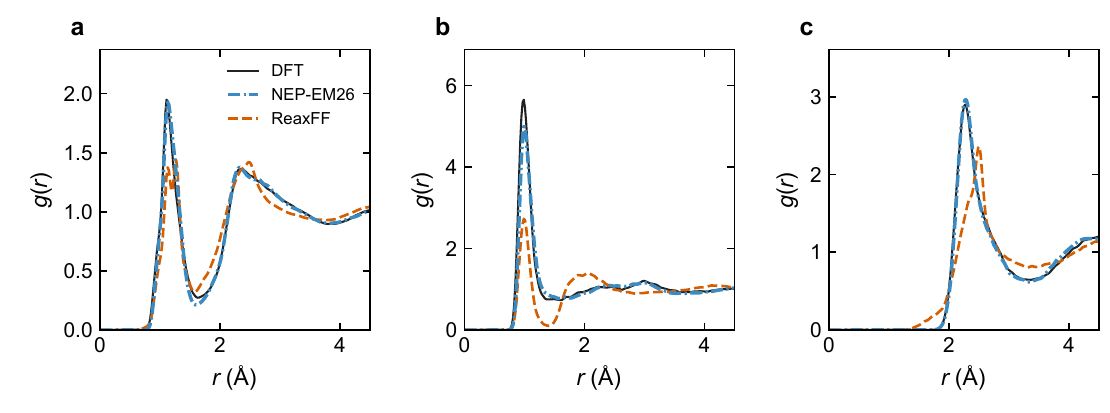}
\caption{CL-20 radial distribution functions (RDFs) from DFT, NEP-EM26, and ReaxFF.
\textbf{a}, Pair-count-weighted composite of the ten C/H/N/O partial RDFs;
\textbf{b}, H--O RDF; \textbf{c}, O--O RDF.
Curves are averaged over an approximately 5\,ps late-time window and Gaussian-broadened
($\sigma_\mathrm{RDF}=0.03$\,\AA{}) for visualization.}
\label{fig:rdf_product}
\end{figure*}

For CL-20, the pair-count-weighted composite of all ten C/H/N/O partial
radial distribution functions (RDFs) from NEP-EM26 retains the principal DFT peak pattern
(Fig.~\ref{fig:rdf_product}a). NEP-EM26 closely follows the first peak and
the broader second peak, whereas ReaxFF underestimates the first peak.
The individual H--O and O--O RDFs show the same contrast (Fig.~\ref{fig:rdf_product}b,c): NEP-EM26 more closely
reproduces their DFT peak shapes, with a mean normalized integrated
absolute deviation of 0.064, compared with 0.278 for ReaxFF. Not all pairs
show the same ordering: NEP-EM26 does not reproduce the short-range C--C peak
(Supplementary Fig.~S3g), and ReaxFF gives the smaller H--H deviation
(0.181 versus 0.221). Supplementary Figs.~S3 and S4 give
the full ten-pair comparisons for CL-20 and RDX, respectively.

\setcounter{dbltopnumber}{1}
\begin{figure*}[!htp]
	\centering
	\includegraphics[width=\textwidth]{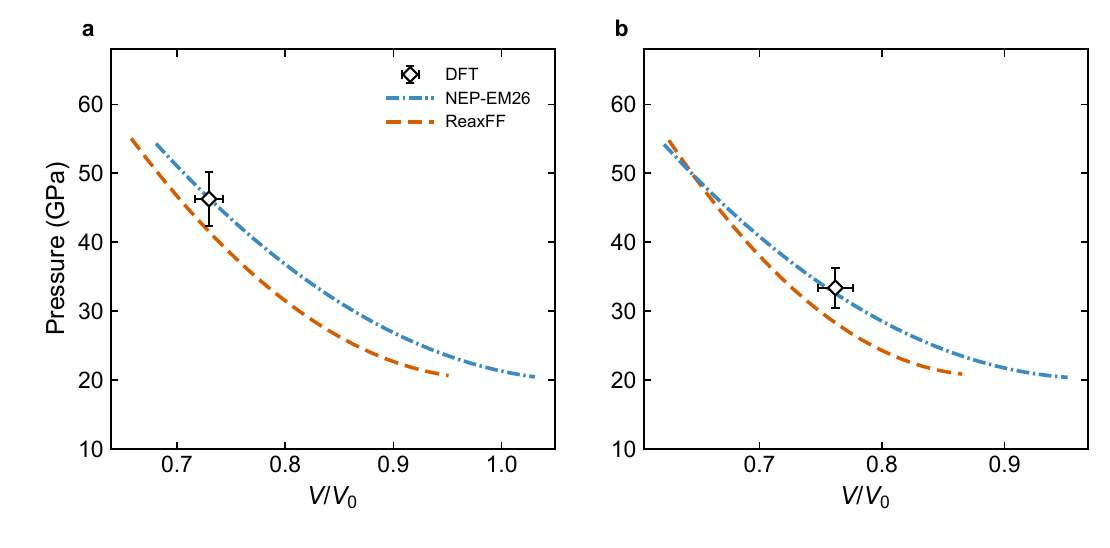}
	\caption{Pressure--volume comparisons for \textbf{a}, CL-20, and \textbf{b}, RDX.
		Curves show quadratic fits to eight NEP-EM26 or ReaxFF trajectory means within the sampled volume range.
		Markers show DFT trajectory means, which are not used in the fits.
        The DFT calculations include the D3(BJ) dispersion correction.
		Error bars indicate temporal standard deviations within each DFT trajectory.
		For each material, model and DFT volumes are normalized by the same model reference volume $V_0$ (Supplementary Information Section~S3.1).}
	\label{fig:cl20_rdx_hugoniot}
\end{figure*}

Averaged over the six C/N/O pairs, the deviation is 0.101 for NEP-EM26 and
0.290 for ReaxFF in CL-20, and 0.168 and 0.311, respectively, in RDX, so
NEP-EM26 gives the smaller deviation in both materials.
Definitions of the deviation metric and pair weighting are given in
Supplementary Information Section~S2.2.

Potential-energy evolution follows the same trend (Supplementary Fig.~S5).
Over the common late-time window of approximately 16--20\,ps, mean
potential-energy changes in CL-20 are 0.649, 0.619, and
0.527\,eV$\cdot$atom$^{-1}$ for DFT, NEP-EM26, and ReaxFF, respectively.
In RDX, the corresponding values are 0.500, 0.442, and 0.357\,eV$\cdot$atom$^{-1}$.

\subsection{Pressure--Volume Response and Detonation Velocities}
\label{sec:detonation}

Figure~\ref{fig:cl20_rdx_hugoniot}
compares the NEP-EM26 and ReaxFF responses of CL-20 and RDX with DFT. Each model curve is fitted
to eight trajectory-averaged states at target pressures from 20 to 55\,GPa
at 5\,GPa intervals.
At the CL-20 DFT mean pressure of 46\,GPa, the fitted volumes differ from
DFT by $+0.10$\% for NEP-EM26 and $-3.73$\% for ReaxFF. For RDX at
33\,GPa, the differences are $-0.91$\% and $-4.65$\%, respectively.
For comparison, the temporal standard deviations of the DFT volumes are 1.8\%
for CL-20 and 1.9\% for RDX. The absolute NEP-EM26 deviations are smaller than the corresponding temporal standard deviations of the DFT volume trajectories at both states, whereas the ReaxFF deviations are larger.

For each material, we sampled states on the Hugoniot of detonation products at eight target pressures
from 20 to 55\,GPa using NEP-EM26 and ReaxFF.
Detonation velocities were then obtained from quadratic fits to these states
using the CJ tangent construction described in Methods
(Table~\ref{tab:detonation}). Because detonation velocity depends on loading
density~\cite{price1983_velocity_density}, we compare the predictions with
literature values at the same densities using the experimental reference
compilation cited in Table~\ref{tab:detonation}.

\begin{table*}[t]
\centering
\begin{threeparttable}
\caption{Detonation-velocity estimates for eight materials using ReaxFF and
NEP-EM26, with literature experimental reference values at the listed densities.
Both models use eight target pressures of 20--55\,GPa and the same quadratic
CJ construction (Supplementary Information Section~S3).
The absolute percentage error is $100\%\cdot\frac{|D-D_\mathrm{ref}|}{D_\mathrm{ref}}$,
where $D$ is the corresponding model estimate and $D_\mathrm{ref}$ is the
literature reference value. Full names of LLM-105,
FOX-7, HMX, and CL-20 are given in Table~\ref{tab:source_errors}; PETN is
pentaerythritol tetranitrate, TNT is 2,4,6-trinitrotoluene, RDX is
1,3,5-trinitro-1,3,5-triazinane, and Tetryl is
$N$-methyl-$N$,2,4,6-tetranitroaniline.
}
\label{tab:detonation}
\renewcommand{\arraystretch}{1.06}
\begin{tabular*}{\textwidth}{@{\extracolsep{\fill}}ccccccc@{}}
\toprule
\multirow{2}{*}{Material} &
\multirow{2}{*}{\shortstack{Density\\(g$\cdot$cm$^{-3}$)}} &
\multicolumn{3}{c}{Detonation velocity (km$\cdot$s$^{-1}$)} &
\multicolumn{2}{c}{Absolute percentage error (\%)} \\
\cmidrule(lr){3-5}\cmidrule(lr){6-7}
& & Reference & ReaxFF & NEP-EM26 & ReaxFF & NEP-EM26 \\
\midrule
CL-20~\cite{utkin2022_cl20}       & 1.980 & 9.380 & 8.487 & 9.057 & 9.5 & 3.4 \\
FOX-7~\cite{trzcinski2016_fox7}   & 1.780 & 8.325 & 7.832 & 8.443 & 5.9 & 1.4 \\
HMX~\cite{suceska2025_exp6,hobbs1993_bkw}\tnote{a}   & 1.890 & 9.115 & 8.593 & 8.942 & 5.7 & 1.9 \\
LLM-105~\cite{li2008_llm105}\tnote{b}      & 1.845 & 7.991 & 7.726 & 8.343 & 3.3 & 4.4 \\
PETN~\cite{suceska2025_exp6,hobbs1993_bkw}\tnote{a}  & 1.763 & 8.270 & 7.358 & 8.423 & 11.0 & 1.9 \\
RDX~\cite{suceska2025_exp6,hobbs1993_bkw}\tnote{a}     & 1.800 & 8.754 & 8.149 & 8.616 & 6.9 & 1.6 \\
TNT~\cite{suceska2025_exp6,hobbs1993_bkw}\tnote{a}         & 1.640 & 6.942 & 6.580 & 7.261 & 5.2 & 4.6 \\
Tetryl~\cite{suceska2025_exp6,hobbs1993_bkw}\tnote{a} & 1.680 & 7.500 & 6.650 & 7.476 & 11.3 & 0.3 \\
\midrule
\multicolumn{5}{l}{Mean absolute percentage error\tnote{c}} & 7.4 & 2.4 \\
\bottomrule
\end{tabular*}
\begin{tablenotes}
\footnotesize
\item[a] These values are tabulated as experimental detonation velocities in
Table~4 of Su\'ceska~\cite{suceska2025_exp6}, where they are attributed to the
Hobbs--Baer compilation~\cite{hobbs1993_bkw}.
\item[b] Measurement for a 95/5 LLM-105/F2314 formulation at 96.4\% of
theoretical maximum density (TMD), not for neat LLM-105.
\item[c] Averaged over the same eight materials for both models, using
unrounded detonation velocities.
\end{tablenotes}
\end{threeparttable}
\end{table*}

For the eight materials in Table~\ref{tab:detonation}, NEP-EM26 gives a mean
absolute percentage error (MAPE) of 2.4\%; individual errors range from
0.3\% to 4.6\%. The MAPE is 2.8\% for the four reactive-MD source materials
and 2.1\% for the four materials without dedicated reactive-MD training
configurations, so the two groups show comparable accuracy. Moving-block bootstrap resampling of the trajectory tails gives a 95\%
sampling interval of 2.35--2.53\% for the eight-material MAPE (Supplementary Information
Section~S3.4 and Tables~S6--S8); this interval covers finite-time sampling
only. The LLM-105 reference refers to a 95/5 LLM-105/F2314 formulation
(Table~\ref{tab:detonation}, note b); excluding it gives MAPEs of 2.2\% for
NEP-EM26 and 8.0\% for ReaxFF.

\begin{figure}[!t]
\centering
\includegraphics[width=\columnwidth]{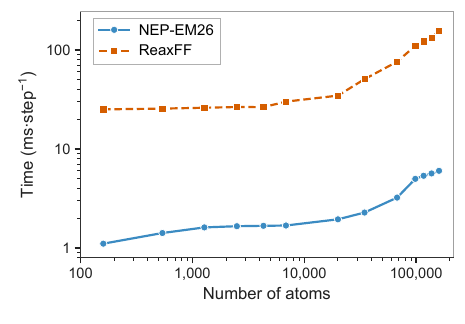}
\caption{GPU performance comparison of NEP-EM26 and ReaxFF. Mean time per molecular dynamics step is plotted against system size for NEP-EM26 in GPUMD and ReaxFF in LAMMPS, measured on the same GPU.
Benchmark details are provided in Supplementary Information Section~S4.}
\label{fig:efficiency}
\end{figure}

For the same eight materials, ReaxFF gives a MAPE of 7.4\%, and NEP-EM26
has smaller errors for seven of them.
ReaxFF underestimates
all eight velocities, whereas NEP-EM26 shows no consistent tendency
toward over- or underestimation. 
To evaluate computational efficiency, we benchmarked NEP-EM26 in GPUMD against GPU-accelerated ReaxFF in LAMMPS on the same GPU (Fig.~\ref{fig:efficiency}). Across the tested system sizes of 160--160{,}000 atoms, both methods exhibit relatively weak system-size dependence for small systems, followed by increasing computational costs at larger sizes. For all tested system sizes, NEP-EM26 runs 15.8--26.0 times faster per MD step than ReaxFF. In practice, this efficiency enables a full eight-pressure D-Hug detonation-velocity calculation for an energetic material to be completed within 30 minutes on the benchmarked GPU.

%%=============================================================
\section{Discussion}
\label{sec:discussion}
%%=============================================================
NEP-EM26 demonstrates that a single DFT-trained machine-learning interatomic potential can enable accurate and efficient detonation-velocity predictions across diverse CHNO energetic materials. Comparisons with DFT show that NEP-EM26 reproduces static energy--volume trends of eight crystals and captures selected structural and pressure--volume features of high-pressure reactive states. For eight explosives, NEP-EM26 achieves a detonation-velocity mean absolute percentage error of 2.4\%, approximately one-third of the 7.4\% obtained with ReaxFF. In the computational efficiency test, NEP-EM26 in GPUMD is more than an order of magnitude faster per molecular dynamics step than ReaxFF in LAMMPS on the same GPU. This combination of predictive accuracy and computational efficiency makes repeated Hugoniostat simulations practical across multiple energetic materials.

The performance of NEP-EM26 is likely supported by the diversity of its training data, which combines molecular bonding and crystal packing environments from CSD structures with compressed and reactive configurations from eleven reactive-MD systems. This complementary coverage enables a single potential to describe both molecular crystals and high-pressure reaction products, including four benchmark materials without dedicated reactive-MD training configurations. Descriptor-space sampling, explicit force and stress fitting, and the short-range repulsive correction may further contribute to model robustness, although their individual effects have not been isolated. The relatively large errors for the HMX and DNAN subsets and the strongly compressed configurations of TNT, Tetryl, and PETN suggest that targeted refinement of these environments could further improve model accuracy.

Building on previous machine-learning potentials for energetic materials~\cite{deepmd_cj2023,nto_dnnp_2024,yoo2021nnff,hamilton2023_nnrf,wen2024_hem_nnp,emff2025,li2025chon_reactive,pham2026_shocked_mlff} and universal neuroevolution potentials~\cite{liangNEP89UniversalNeuroevolution2026}, NEP-EM26 demonstrates the value of combining a GPU-efficient interatomic potential with multilevel validation extending from DFT-based atomistic properties to experimentally measured detonation velocities. Its ability to predict materials without dedicated reactive-MD training configurations highlights the potential for cross-material applications while maintaining substantially lower per-step computational cost than the tested ReaxFF baseline.

The scope of these findings should nevertheless be considered within the adopted D-Hug framework. D-Hug samples product-like states rather than explicitly resolving detonation-front propagation or initiation kinetics. The bootstrap intervals quantify only finite-time sampling within the present trajectories; finite-size effects and variation across independent initial conditions have not been quantified.
The LLM-105 experimental reference also corresponds to a formulation rather than the neat material. Further assessment of reference-state consistency, sampling convergence, and reaction-product chemistry will be important for extending the predictive reliability of this approach to broader loading conditions and energetic-material systems.

More broadly, this approach provides a practical route for extending first-principles-based detonation-performance predictions to a wider range of energetic materials. Future efforts could prioritize expanding coverage of high-pressure reactive environments, improving descriptions of reaction-product chemistry, and developing more robust strategies for assessing prediction uncertainties. Such advances would further improve the accuracy and transferability of machine-learning potentials and enable more reliable predictions of macroscopic energetic-material performance.

%%=============================================================
\section{Methods}
\label{sec:methods}
%%=============================================================

\subsection{NEP-EM26 Model and Optimization}
\label{sec:nep}

NEP-EM26 uses the NEP4 formulation implemented in
GPUMD~\cite{fan2022gpumd,gpumd4}. Radial and angular descriptors encode
neighbor distances and many-body geometry~\cite{fan2021nep,fan2022_nep_multicomponent}
and are mapped to atomic site energies by element-specific feed-forward
networks~\cite{song2024_nep4}. Based on preliminary tests of cutoff
pairs and hidden-layer widths, we chose radial and angular cutoffs of
5.0 and 4.0\,\AA{}, respectively, and a single hidden layer of
70 neurons with hyperbolic-tangent activation in each network. A universal Ziegler--Biersack--Littmark (ZBL)
correction~\cite{biersack1982_zbl} supplies short-range repulsion, with inner
and outer switching radii of 1.0 and 2.0\,\AA{}, respectively.

The model parameters were trained from random initialization with the separable natural evolution
strategy~\cite{schaul2011snes} against PBE-D3(0) energy, force, and stress labels in two consecutive stages.
The first stage used mini-batch optimization with loss weights of $\lambda_E:\lambda_F:\lambda_V=1.0:1.0:0.1$ for the energy, force, and stress-derived virial terms.
The second stage continued from the first-stage parameters with full-batch optimization over the training set and loss weights of $\lambda_E:\lambda_F:\lambda_V=5.0:1.0:1.0$. Both stages used a population size of 50.
Complete input and training settings are given in the Supplementary
Information Section~S5.1.

\subsection{Training Dataset and DFT Reference Calculations}
\label{sec:data}

The dataset combines crystal structures from the Cambridge Structural
Database (CSD) with configurations sampled from reactive molecular dynamics
(reactive-MD) trajectories.

\textbf{CSD static structures.}
We queried the Cambridge Structural Database~\cite{groom2016csd} for
experimentally determined crystal structures containing only C, H, N,
and O. We then removed structures with more than 500 atoms, near-duplicate structures, and structures outside the adopted distance and density ranges. After label-quality screening,
54{,}574 CHNO crystal structures remained.

\begin{figure*}[!t]
\centering
\includegraphics[width=\textwidth]{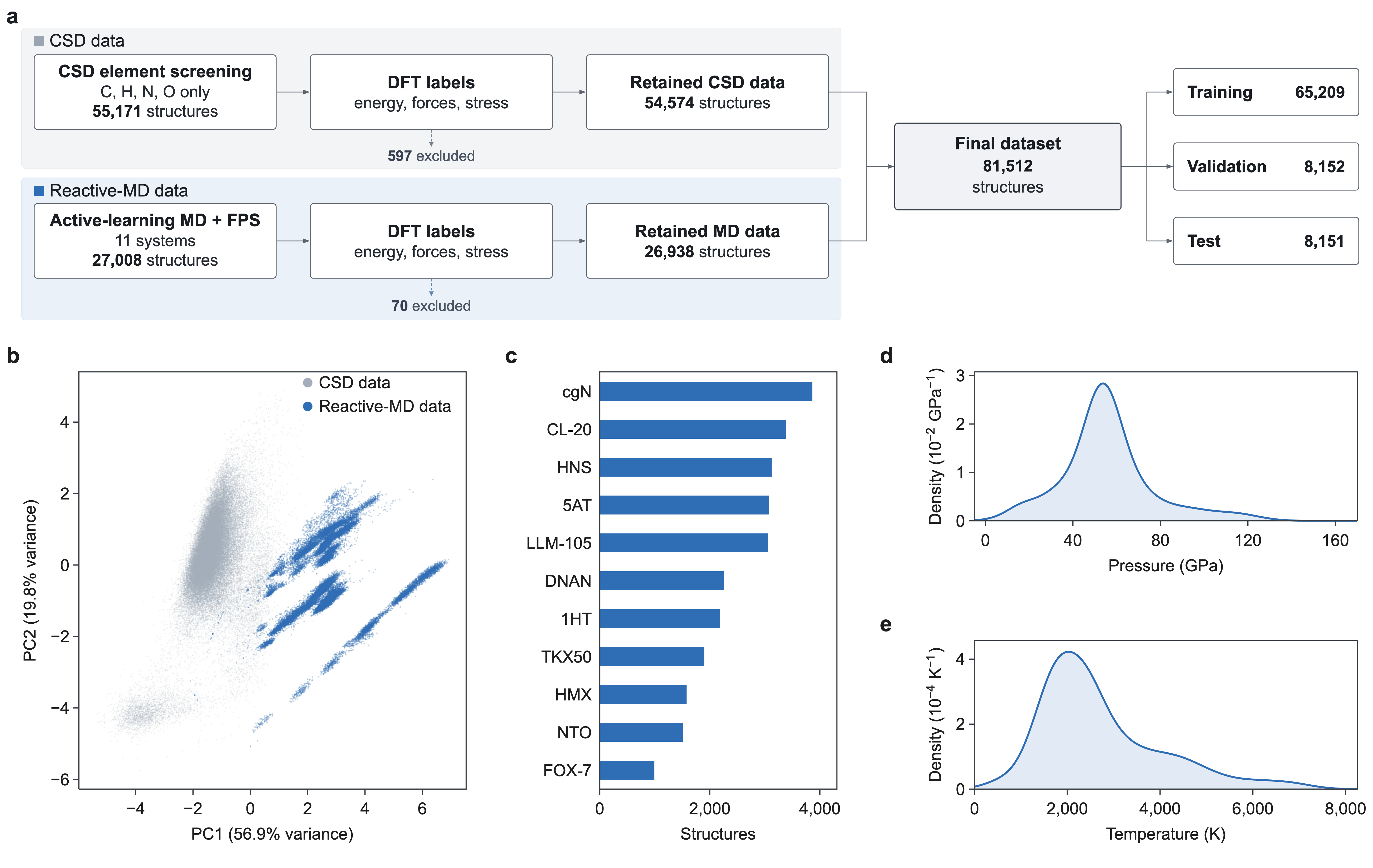}
\caption{Construction and composition of the NEP-EM26 dataset.
\textbf{a}, Dataset construction from CSD structures and reactive-MD configurations; 597 CSD and 70 reactive-MD candidates were excluded due to unphysical structures, anomalous DFT labels, or failed single-point calculations (Supplementary Information Section~S5.2).
\textbf{b}, Principal component projection of 81{,}512 configurations using eight standardized physical and compositional features: mass density, pressure, energy per atom, volume per atom, and C, H, N, and O atomic fractions.
\textbf{c}, Composition of the reactive-MD subset by data source.
\textbf{d,e}, Pressure and temperature distributions of the reactive-MD configurations, shown as probability densities.
}
\label{fig:dataset}
\end{figure*}

\textbf{Reactive-MD trajectories.}
An in-house active-learning workflow collected 182{,}122 candidate
configurations from eleven source systems: cgN, CL-20, HNS, 5AT, LLM-105,
DNAN, 1HT, TKX50, HMX, NTO, and FOX-7 (Fig.~\ref{fig:dataset}). Cubic
gauche nitrogen (cgN) adds a nitrogen-rich source alongside the CHNO energetic
materials. For each configuration, the atomic descriptors of a preliminary NEP were averaged into a structure-level descriptor vector. Farthest-point sampling (FPS)
within each source system, with a minimum descriptor-space distance of 0.007,
retained 27{,}008 configurations. Filtering and sampling details are given in
Supplementary Information Section~S5.2.

Single-point DFT calculations at the PBE-D3(0) level (zero-damping D3)~\cite{kresse1996vasp,perdew1996pbe,grimme2010d3}
provided total energies, atomic forces, and stress tensors without geometry
optimization. The final dataset contains 81{,}512 configurations, including
26{,}938 reactive-MD configurations with complete DFT labels. DFT settings and the dispersion convention are
detailed in
Supplementary Information Section~S5.3.

Figure~\ref{fig:dataset} provides a schematic summary of the dataset sources,
retained counts, and final data splits. After configuration selection and label screening, the dataset was split randomly by configuration into
training (65{,}209), validation (8{,}152), and test (8{,}151) sets, approximately 80:10:10.
Stresses are reported with compression taken as positive and exclude kinetic contributions.

\Needspace{5\baselineskip}
\subsection{Static Energy--Volume Benchmark}
For each of the eight DFT-optimized crystals in Fig.~\ref{fig:static_eos}, we
generated 12 volumes spanning $V/V_0=0.60$ to 1.50, including 1.00, giving 96 configurations in total. Cell
vectors and Cartesian atomic coordinates were scaled by the same factor,
keeping fractional coordinates fixed. DFT, NEP-EM26, and ReaxFF single-point
energies were evaluated on identical geometries without further relaxation.
Each energy curve was shifted to zero at $V/V_0=1.00$, and the reported RMSE measures the deviation of the model $\Delta E$ from DFT over all 96 configurations. DFT settings are
given in Supplementary Information Section~S1.3.

\subsection{Reactive Dynamics and Detonation-Velocity Calculations}
\label{sec:nphug}

\textbf{CL-20 and RDX dynamics benchmarks.}
\label{sec:rdf_methods}
The benchmarks use 144-atom CL-20 and 168-atom RDX cells. One DFT
trajectory per material was computed with PBE and D3(BJ) dispersion~\cite{grimme2011_d3bj}
under isotropic constant-pressure Hugoniostat (NPHug) dynamics, spanning approximately 25\,ps for CL-20 and 20\,ps for RDX. NEP-EM26 and
ReaxFF started from the DFT initial cells and coordinates, with
target pressures set to the late-time DFT mean pressures of 46 and 33\,GPa, respectively.
Each model trajectory ran for 40\,ps with a time step of 0.4\,fs and 300\,K
initial velocities. Model-specific reference energies were taken from 4\,ps
constant-volume (NVT) trajectories of the original crystals.
Reference-state and initialization settings are detailed in Supplementary
Information Section~S2.1.

RDFs were averaged with equal frame weights over the same late-time windows as the DFT trajectories
and normalized by instantaneous volume. Agreement with DFT was quantified
as the normalized integrated absolute
deviation of the unsmoothed RDFs. Potential-energy changes were expressed per
atom relative to each method's first common saved frame.
Energy traces covered approximately 20\,ps for both materials.
Sampling and normalization
details are given in Supplementary Information Section~S2.

\textbf{D-Hug framework.}
We coupled NEP-EM26 to the
Decoupled Hugoniostat (D-Hug) framework, which samples 
states on the Hugoniot of detonation products from stoichiometry-preserving, product-like
pseudo-initial configurations and determines the CJ state
from a tangent Rayleigh line. The GPUMD NPHug implementation~\cite{pan2024_gpumd_nphug} adapts the
formulations of Maillet et al. and Ravelo et al.~\cite{maillet2000nphug,ravelo2004nphug}. The construction of the pseudo-initial cells and the Hugoniot reference states is described in Supplementary Information Section~S3.

For the CJ construction, the Rayleigh line through $(x,P)=(1,0)$ is tangent
to the fitted Hugoniot of detonation products at the CJ state. Here $P$ is pressure and
$x=V/V_\mathrm{load}$ is cell volume $V$ normalized by the loading reference
volume $V_\mathrm{load}$. Its slope, $m<0$, gives the detonation velocity
$D=\sqrt{-m/\rho_0}$, where $\rho_0$ is the loading density.

For NEP-EM26, eight isotropic NPHug trajectories per material were run in GPUMD from 20 to 55\,GPa
at 5\,GPa intervals, each for 40\,ps with a time step of 0.4\,fs, a
pressure-coupling period of 2{,}000 steps, and initial velocities drawn at 1{,}000\,K.
Pressure and volume were averaged over the final 20\% of each trajectory.
The state means were fitted by unweighted ordinary least squares to
$P(x)=ax^2+bx+c$, where $a$, $b$, and $c$ are fitted coefficients with units
of pressure. The CJ pressure and detonation velocity were obtained from the
tangent construction above.

ReaxFF used the same initial product cells and atomic coordinates, eight
target pressures, loading densities, and quadratic CJ construction.
A CJ state was reported only if it had positive pressure and
$0<x<1$ within the sampled reduced-volume range. Method-specific reference preparation and trajectory settings are
given in Supplementary Information Section~S3.3.

\textbf{CL-20 and RDX pressure--volume comparisons.}
\label{sec:cl20_rdx_hugoniot_methods}
For Fig.~\ref{fig:cl20_rdx_hugoniot}, CL-20 and RDX states were sampled at eight target pressures from 20
to 55\,GPa at 5\,GPa intervals; each trajectory ran for 40\,ps, and pressure and volume
were averaged over its final 20\%. For each material, both models use the same initial
cell and atomic coordinates, with velocities initialized at 1{,}000\,K.
The same quadratic least-squares procedure was applied to these state means,
with $x=V/V_0$ and a common reference volume $V_0$ for each panel.
Initialization and reference volumes are detailed in Supplementary Information
Section~S3.1. The influence of the D3(0) and D3(BJ) dispersion conventions
on the NEP-EM26 pressure--volume curves is examined separately in
Supplementary Information Section~S3.2.

\Needspace{4\baselineskip}
\textbf{\mbox{Computational-efficiency benchmark.}}
\label{sec:efficiency_methods}
The NEP-EM26--ReaxFF timing comparison used GPUMD
5.5~\cite{fan2022gpumd,gpumd4} and the KOKKOS-accelerated ReaxFF
implementation in LAMMPS~\cite{thompson2022_lammps}.
Both were timed on the same NVIDIA RTX PRO 6000 Blackwell Server Edition
GPU with no other jobs running, and the ReaxFF timings include
charge equilibration~\cite{rappe1991_qeq} at every step.
Timing results are shown in Fig.~\ref{fig:efficiency}, with detailed settings and
statistics provided in Supplementary Information Section~S4.

\raggedbottom
\backmatter

\pdfbookmark[1]{Data availability}{data-availability}
\bmhead*{Data availability}

The DFT training, validation, and test dataset generated in this study is not publicly deposited and is available from the corresponding authors via email upon reasonable request. Structures derived from the Cambridge Structural Database (CSD) are subject to the CSD license terms. All other data supporting the findings of this study are available within the article and its Supplementary Information.

\Needspace{5\baselineskip}
\pdfbookmark[1]{Code availability}{code-availability}
\bmhead*{Code availability}

NEP-EM26 was trained and run with the open-source GPUMD package~\cite{fan2022gpumd,gpumd4}. The trained NEP-EM26 potential and the custom scripts used for simulation workflows, data processing, analysis, and figure generation are publicly available at \url{https://github.com/LuckRookie/CNOH_NEP}.

%%=============================================================
%% Bibliography
%%=============================================================

% Shared database; sn-jnl selects sn-nature.bst via the class option.
\begingroup
\urlstyle{same}
\renewcommand{\bibfont}{\fontsize{9}{10.5}\selectfont}
\setlength{\bibsep}{1pt}
\apptocmd{\thebibliography}{\clubpenalty=10000 \widowpenalty=10000}{}{}
\bibliography{references}
\endgroup

\pdfbookmark[1]{Acknowledgements}{acknowledgements}
\bmhead*{Acknowledgements}
This work was supported by the Advanced Materials-National Science and Technology Major Project (No. 2024ZD0606900).

\pdfbookmark[1]{Author contributions}{author-contributions}
\bmhead*{Author contributions}

Z.Q. and J.D. performed the main computational work, including dataset
curation, DFT reference calculations, NEP-EM26 training, molecular-dynamics
simulations, and data analysis. Z.L. constructed and provided the reactive-MD
dataset. Z.F., X.S., and W.M. supervised the study and analyzed data. Z.Q.
and J.D. drafted the manuscript, and all authors reviewed and revised the
manuscript and approved the final version.

\pdfbookmark[1]{Competing interests}{competing-interests}
\bmhead*{Competing interests}

All authors declare no financial or non-financial competing interests.

\enlargethispage{4\baselineskip}
\pdfbookmark[1]{Supplementary information}{supplementary-information}
\bmhead*{Supplementary information}

Supplementary information accompanies this manuscript and provides dataset-screening details, NEP-EM26 and DFT settings, benchmark protocols, and
material-specific detonation-velocity calculation details.

\end{document}

\title{Supplementary Information for ``NEP-EM26: A Neuroevolution Potential for
Accurate and Efficient Detonation Velocity Prediction in CHNO Energetic Materials''}

\author[1]{\fnm{Zhenkun} \sur{Qiu}}
\author[1]{\fnm{Junwen} \sur{Duan}}
\author[1]{\fnm{Zifeng} \sur{Liu}}
\author*[2]{\fnm{Zheyong} \sur{Fan}}\email{brucenju@gmail.com}
\author*[1,3,4]{\fnm{Xuecheng} \sur{Shao}}\email{shaoxc@jlu.edu.cn}
\author*[1,3,4]{\fnm{Wenhui} \sur{Mi}}\email{mwh@jlu.edu.cn}

\affil[1]{
  \orgdiv{Key Laboratory of Material Simulation Methods \& Software of Ministry of Education, College of Physics},
  \orgname{Jilin University},
  \orgaddress{
    \city{Changchun},
    \postcode{130012},
    \country{China}}}

\affil[2]{
  \orgdiv{College of Physical Science and Technology},
  \orgname{Bohai University},
  \orgaddress{
    \city{Jinzhou},
    \postcode{121013},
    \country{China}}}

\affil[3]{
  \orgdiv{State Key Laboratory of High Pressure and Superhard Materials, College of Physics},
  \orgname{Jilin University},
  \orgaddress{
    \city{Changchun},
    \postcode{130012},
    \country{China}}}

\affil[4]{
  \orgdiv{International Center of Future Science},
  \orgname{Jilin University},
  \orgaddress{
    \city{Changchun},
    \postcode{130012},
    \country{China}}}

\maketitle

\section{Held-out Accuracy and Static Energy--Volume Benchmarks}

\subsection{Held-out mean absolute errors and force scale}

On the 8{,}151-configuration held-out test set, the pooled mean absolute errors
(MAEs) are 14.9\,meV$\cdot$atom$^{-1}$ for energy, 400\,meV$\cdot$\AA$^{-1}$ for
Cartesian force components, and 0.75\,GPa for stress components.
Table~\ref{tab:source_mae} gives the corresponding source-resolved values for
Cambridge Structural Database (CSD) structures and reactive-MD configurations.
As in Fig.~1 and the source-resolved root-mean-square error (RMSE) table in the
main article, energies are normalized
by atom count, force components are pooled over atoms and Cartesian directions,
and stress errors are pooled over the six independent tensor components.
For each configuration, the native per-atom tensor output is multiplied by
its atom count and divided by its cell volume to obtain
compression-positive stress, using
$1\,\mathrm{eV}\cdot\text{\AA}^{-3}\approx160.218\,\mathrm{GPa}$.
No kinetic contribution is included in this configuration-level comparison.

To place the force RMSE on the scale of the sampled forces, the root mean
square of the 3{,}095{,}166 DFT Cartesian force components is
$3.40\times10^3$\,meV$\cdot$\AA$^{-1}$. The pooled force RMSE is 18.8\% of this
reference RMS magnitude, using the same components as Fig.~1.

\begin{sitable}
\caption{Source-resolved mean absolute errors on the held-out test set.}
\label{tab:source_mae}
\begin{tabular*}{\sitablewidth}{@{\extracolsep{\fill}}cccc}
\toprule
Source & \shortstack{Energy per atom\\(meV$\cdot$atom$^{-1}$)} &
\shortstack{Force component\\(meV$\cdot$\AA$^{-1}$)} &
\shortstack{Stress component\\(GPa)} \\
\midrule
CSD    & 13.3 & 195 & 0.47 \\
cgN    & 15.6 & 711 & 1.34 \\
CL-20   & 16.2 & 573 & 0.99 \\
HNS    & 17.5 & 680 & 1.39 \\
5AT    & 19.3 & 631 & 1.37 \\
LLM-105 & 16.0 & 602 & 1.25 \\
DNAN   & 23.9 & 738 & 2.00 \\
1HT    & 17.9 & 691 & 1.52 \\
TKX50  & 16.6 & 516 & 1.07 \\
HMX    & 26.0 & 580 & 1.29 \\
NTO    & 15.8 & 574 & 1.09 \\
FOX-7   & 24.7 & 616 & 1.41 \\
\bottomrule
\end{tabular*}
\end{sitable}

\subsection{Temperature--pressure resolved held-out errors}

\begin{figure}[!htbp]
\centering
\includegraphics[width=\textwidth]{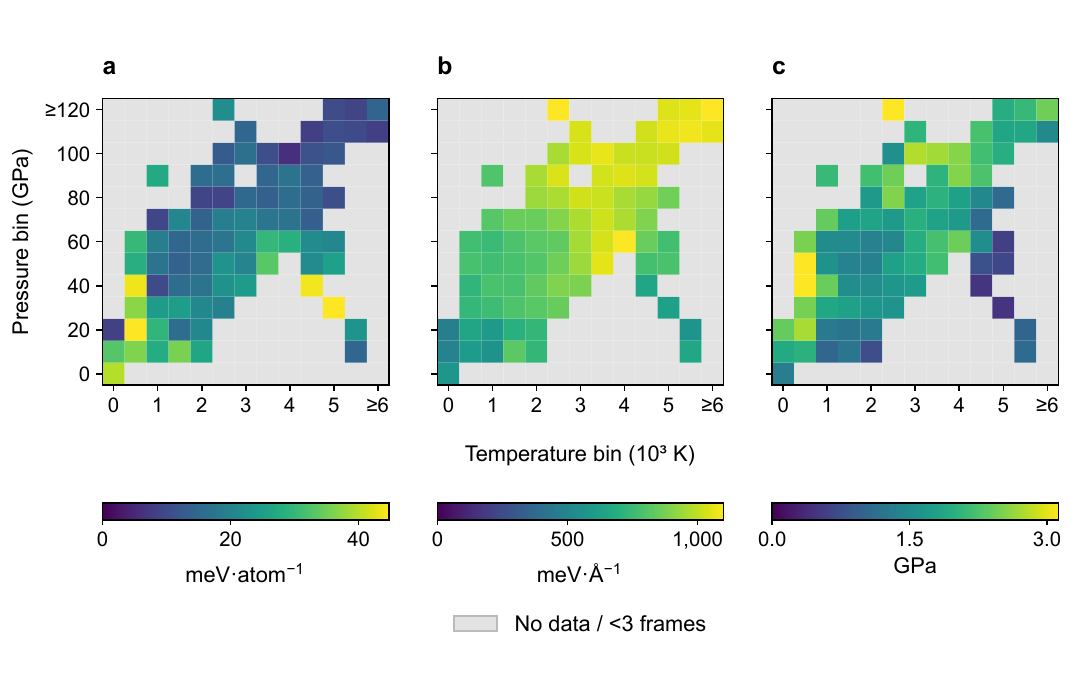}
\caption{Temperature--pressure distribution of held-out errors for the
reactive-MD portion of the PBE-D3(0)-labeled test set.
\textbf{a}, Mean absolute energy error per atom;
\textbf{b}, mean force-component RMSE;
\textbf{c}, mean stress-component RMSE in GPa, for each temperature--pressure cell.
The physical grid uses
500\,K and 10\,GPa bins; labels are thinned to 1{,}000\,K and 20\,GPa, and the
last cells denote $\geq6{,}000$\,K and $\geq120$\,GPa. Gray cells contain no data or
fewer than three frames.}
\label{fig:supp_temp_pressure}
\end{figure}

Figure~\ref{fig:supp_temp_pressure} resolves the held-out errors over the
temperature and pressure metadata available for the reactive-MD portion of
the PBE-D3(0)-labeled test set, where D3(0) denotes zero-damping dispersion.
The test set contains 8{,}151 configurations, of which
5{,}503 are CSD structures without trajectory thermodynamic
metadata and 2{,}648 are reactive-MD configurations. Thermodynamic records were
matched by source trajectory and frame identifier. Inclusion required an
available source timestamp and agreement within 0.1\,GPa between the
source-trajectory pressure and the matched thermodynamic-log pressure.
These criteria retained 2{,}637 frames. The remaining 11 frames did not satisfy
the timestamp/pressure-matching criteria. They were excluded
only from this temperature--pressure map and remain in the overall test-set
error statistics.

The map uses 500\,K temperature bins and 10\,GPa pressure bins, with
the final bins combining values at or above 6{,}000\,K and 120\,GPa, respectively.
For readability, only every 1{,}000\,K and 20\,GPa tick is labeled. Cell values are means over the frames in each
bin, and cells containing fewer than three frames are shown in gray. Energy is
reported as the absolute per-atom energy error, while force and stress are
reported as frame-level RMSEs over their corresponding reported components.

\subsection{Static energy--volume paths}

\begin{figure}[!htbp]
\centering
\includegraphics[width=\textwidth]{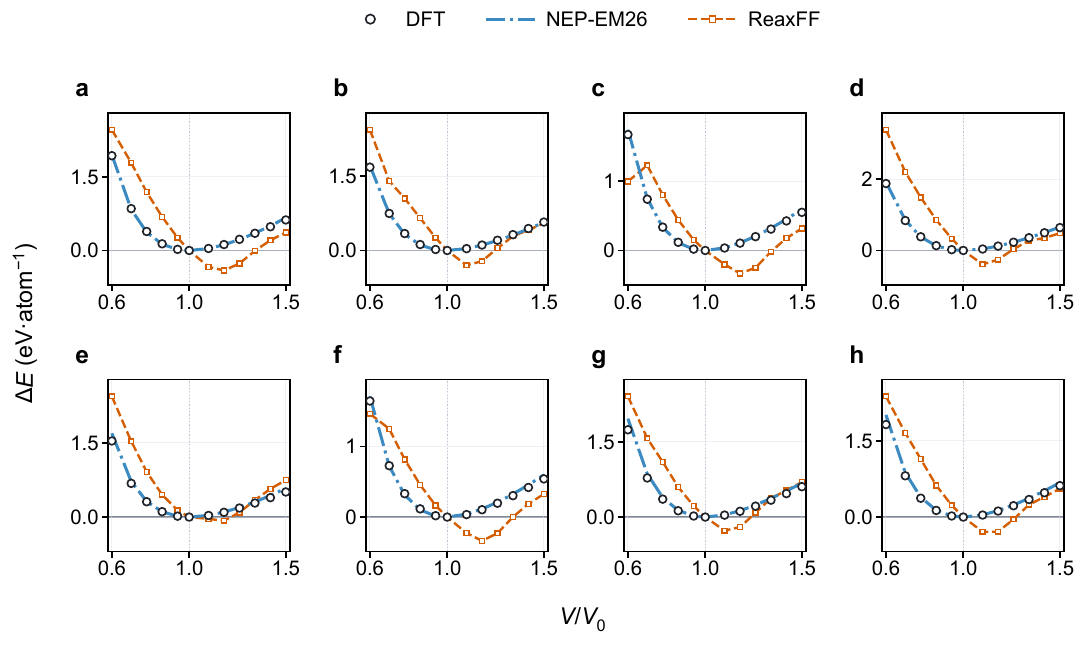}
\caption{Static energy--volume comparison with the earlier Strachan
parameterization~\cite{strachan2003rdx}. Relative energy per atom $\Delta E$ is plotted against normalized
volume $V/V_0$ for \textbf{a}, CL-20; \textbf{b}, FOX-7; \textbf{c}, HMX;
\textbf{d}, LLM-105; \textbf{e}, PETN; \textbf{f}, RDX; \textbf{g}, TNT;
and \textbf{h}, Tetryl.
Within each panel, the DFT reference points, NEP-EM26 curve, and ReaxFF curve
share the same energy scale, and each curve is independently aligned to zero at
$V/V_0=1.00$. This display complements the ReaxFF comparison in Fig.~2 of the
main article.}
\label{fig:supp_eos_strachan}
\end{figure}

The static DFT calculations were performed with VASP 6.3.0 and 6.5.1,
using PAW-PBE datasets and zero-damping D3(0) (\texttt{IVDW}=11).
Final reference-cell optimizations relaxed the atomic coordinates and cell
(\texttt{IBRION}=2, \texttt{ISIF}=3), with a 700\,eV cutoff,
\texttt{KSPACING}=0.7\,\AA$^{-1}$, an electronic threshold of
$10^{-6}$\,eV, and a force convergence criterion of 0.01\,eV$\cdot$\AA$^{-1}$.
The fixed-geometry EOS calculations retained the cutoff, spacing, and
electronic threshold, with up to 200 electronic iterations and either
\texttt{ALGO=Fast} or \texttt{ALGO=Normal}. EOS energies were taken from the zero-smearing extrapolated
\texttt{energy(sigma->0)} values. Reference-label settings are listed in
Section~S5.3.
The DFT-optimized CL-20, FOX-7, HMX, LLM-105, PETN, RDX, TNT, and Tetryl
reference structures were used as the starting geometries. The cell vectors
and Cartesian atomic coordinates were scaled by the same factor, keeping
fractional coordinates fixed. DFT, NEP-EM26, and the Strachan~\cite{strachan2003rdx} (Fig.~\ref{fig:supp_eos_strachan})
and Liu--Goddard~\cite{liu2011_reaxfflg} ReaxFF parameterizations
were evaluated on the same configurations using single-point energy
calculations, without further relaxation of the atomic positions or cell. The same eight
materials are used in the detonation-velocity benchmark. The twelve volume
ratios were 0.60, 0.70, 0.78, 0.86, 0.94,
1.00, 1.10, 1.18, 1.26, 1.34, 1.42, and 1.50. Each method was shifted
independently to zero energy at
$V/V_0=1.00$, where $V$ is the cell volume and $V_0$ is the DFT-optimized
reference volume. The DFT and
NEP-EM26 curves therefore share the PBE-D3(0) convention, whereas each ReaxFF curve
retains the energy convention of its parameterization.

The RMSEs and MAEs of the independently aligned energy--volume curves are listed in
Table~\ref{tab:eos_mae}. Both metrics use the same 96 matched volume
points as the RMSE analysis in the main article; MAEs provide a
complementary measure less sensitive to the largest residuals.
The two material groups distinguish the four crystals with dedicated reactive-MD
configurations (CL-20, FOX-7, HMX, and LLM-105) from the four without them
(PETN, RDX, TNT, and Tetryl).

\begin{sitable}
\caption{Root-mean-square and mean absolute errors for the independently
aligned static energy--volume curves.}
\label{tab:eos_mae}
\begin{tabular*}{\sitablewidth}{@{\extracolsep{\fill}}ccc}
\toprule
Model and comparison scope & \shortstack{RMSE\\(meV$\cdot$atom$^{-1}$)} & \shortstack{MAE\\(meV$\cdot$atom$^{-1}$)} \\
\midrule
NEP-EM26, all eight materials & 42.1 & 21.6 \\
NEP-EM26, reactive-MD source materials & 17.4 & 10.4 \\
NEP-EM26, materials outside the reactive-MD sources & 57.0 & 32.8 \\
NEP-EM26, CL-20 & 15.7 & 10.8 \\
NEP-EM26, FOX-7 & 11.4 & 8.1 \\
NEP-EM26, HMX & 25.2 & 15.2 \\
NEP-EM26, LLM-105 & 13.9 & 7.5 \\
NEP-EM26, PETN & 52.8 & 30.1 \\
NEP-EM26, RDX & 29.0 & 21.2 \\
NEP-EM26, TNT & 74.4 & 44.9 \\
NEP-EM26, Tetryl & 61.9 & 34.9 \\
Strachan ReaxFF, all eight materials & 469.1 & 360.9 \\
Liu--Goddard ReaxFF, all eight materials & 608.2 & 442.0 \\
\bottomrule
\end{tabular*}
\end{sitable}

\section{Local Structure and Energy Benchmarks}

\subsection{Trajectory sources and matched-pressure calculations}
\label{sec:supp_matched_dynamics}

The detailed comparisons use 144-atom CL-20
(C$_{24}$H$_{24}$N$_{48}$O$_{48}$) and 168-atom RDX
(C$_{24}$H$_{48}$N$_{48}$O$_{48}$) systems. For each material, one pre-existing ABACUS product-Hugoniot trajectory was selected for comparison based on the availability of a suitable reference state near the target pressure. These comparisons therefore represent a single DFT reference state per material.
The CL-20 radial distribution functions (RDFs) are shown in the main article,
with full comparisons for both materials below.

Both selected DFT calculations used ABACUS 3.7.0~\cite{li2016_abacus}
with PBE~\cite{perdew1996pbe}, D3(BJ)~\cite{grimme2011_d3bj}, a numerical
atomic-orbital basis, a 100\,Ry grid cutoff, and $\Gamma$-only sampling.
The orbitals contained two radial functions for each occupied angular channel
and one polarization function (2s2p1d for C, N, O; 2s1p for H), with radii
of 8, 6, 8, and 6\,bohr for C, H, N, and O, respectively. All four elements used SG15 optimized norm-conserving Vanderbilt (ONCV) PBE pseudopotentials. Production calculations used a self-consistent field (SCF)
convergence threshold of $2\times10^{-5}$,
a maximum of 100 iterations, Gaussian smearing of 0.015\,Ry, and Broyden mixing
with coefficient 0.8. The production ensemble was an isotropic constant-pressure
Hugoniostat (NPHug), with a
0.4\,fs time step and five-element temperature and pressure chains. The
starting states were prepared at fixed cell volume, followed by variable-cell
production. The available DFT records span approximately 25.54\,ps for CL-20
and 19.96\,ps for RDX. RDF statistics use the final 20\% of these records;
energy comparisons use the common
20\,ps range specified below.

For each material, NEP-EM26 and Liu--Goddard ReaxFF used the actual DFT
step-zero cell and coordinates.
The model target pressures were the DFT tail-average diagnostic pressures,
46.27\,GPa for CL-20 and 33.37\,GPa for RDX; the nominal DFT input pressures
were 46.28 and 33.36\,GPa. Each model trajectory completed 100{,}000 steps at
0.4\,fs, with coordinates and thermodynamic quantities saved every 8\,fs.
NEP-EM26 used GPUMD 5.5 with isotropic NPHug~\cite{pan2024_gpumd_nphug} and a pressure-coupling period
of 2{,}000 steps. ReaxFF used an isotropic constant-pressure Hugoniostat integrator implemented within ASE~\cite{larsen2017_ase}, with
\texttt{tau\_t}=100 and \texttt{tau\_p}=2000 integration steps
(40 and 800\,fs), coupled to LAMMPS force and
stress evaluations. The Hugoniot energy constraint and reported pressure
include configurational and kinetic contributions; the kinetic stress is
added once to the calculator stress. Temperature uses $3N$ degrees of
freedom, where $N$ is atom count, and the target-temperature floor is 300\,K.
The Liu--Goddard parameterization used \texttt{lgvdw yes};
charge equilibration was applied every step with a $10^{-6}$ tolerance and
a 10\,\AA{} upper cutoff. The NEP-EM26 reference and production trajectories
used the direct learned potential without an additional run-time dispersion
term. The dispersion-convention sensitivity analysis is described in
Section~\ref{sec:supp_d3_sensitivity}.

The Hugoniot reference volumes were 1{,}428\,\AA$^3$ for CL-20 and
1{,}639\,\AA$^3$ for RDX, corresponding to loading densities of
2.038 and 1.800\,g$\cdot$cm$^{-3}$. The unscaled reference-crystal volumes
were 1{,}430 and 1{,}741\,\AA$^3$, and the initial product cell volumes
were 1{,}440 and 1{,}680\,\AA$^3$, respectively.
Each model's reference energy was the mean total energy (potential plus
kinetic) over the final 0.8\,ps of a 4\,ps, 300\,K constant-volume (NVT) trajectory
of the unscaled reference crystal, without preceding relaxation. NEP-EM26
used a Bussi--Donadio--Parrinello thermostat~\cite{bussi2007_rescaling}, and ReaxFF used Langevin dynamics.
Absolute reference energies have method-specific zeros.
The DFT references instead used the final energy from 300\,K NVT
calculations lasting 1\,ps for CL-20 (0.5\,fs steps) and 0.8\,ps for RDX
(0.4\,fs steps), with a one-element Nos\'e--Hoover chain and an SCF
threshold of $2\times10^{-6}$. The Hugoniot reference pressure was
$10^{-8}$\,GPa.

\begin{sitable}
\caption{Common coordinate grids used for RDFs. Each listed
grid is shared by DFT, NEP-EM26, and ReaxFF without interpolation.}
\label{tab:rdf_trajectories}
\begin{tabular*}{\sitablewidth}{@{\extracolsep{\fill}}ccccc}
\toprule
Material & Atoms & Model target (GPa) & Window (ps) & Frames \\
\midrule
CL-20 & 144 & 46.27 & 20.432--25.536 & 639 \\
RDX & 168 & 33.37 & 15.968--19.952 & 499 \\
\bottomrule
\end{tabular*}
\end{sitable}

\subsection{RDF analysis}

\begin{figure}[!htbp]
\centering
\includegraphics[width=\textwidth]{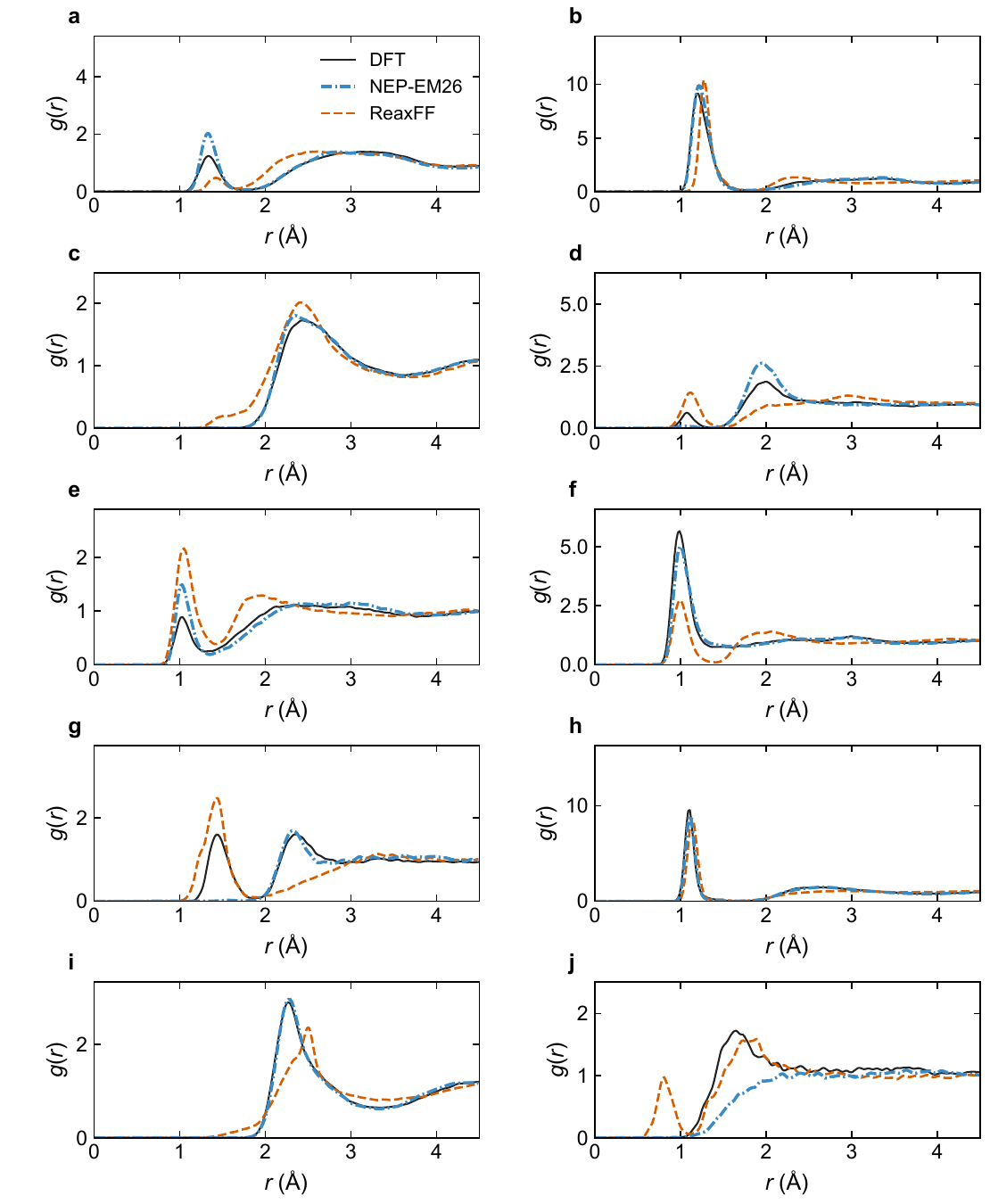}
\caption{All ten C/H/N/O partial RDFs for the selected CL-20 comparison:
\textbf{a}, C--N; \textbf{b}, C--O; \textbf{c}, N--O;
\textbf{d}, C--H; \textbf{e}, H--N; \textbf{f}, H--O;
\textbf{g}, C--C; \textbf{h}, N--N; \textbf{i}, O--O; \textbf{j}, H--H.
H--O and O--O from main Fig.~3b,c are included for completeness.
The analysis uses the same 639 common frames over 20.432--25.536\,ps,
normalization, and display broadening as main Fig.~3.
DFT is black solid, NEP-EM26 blue dash-dotted, and ReaxFF orange dashed.}
\label{fig:supp_cl20_rdf}
\end{figure}

\begin{figure}[!htbp]
\centering
\includegraphics[width=\textwidth]{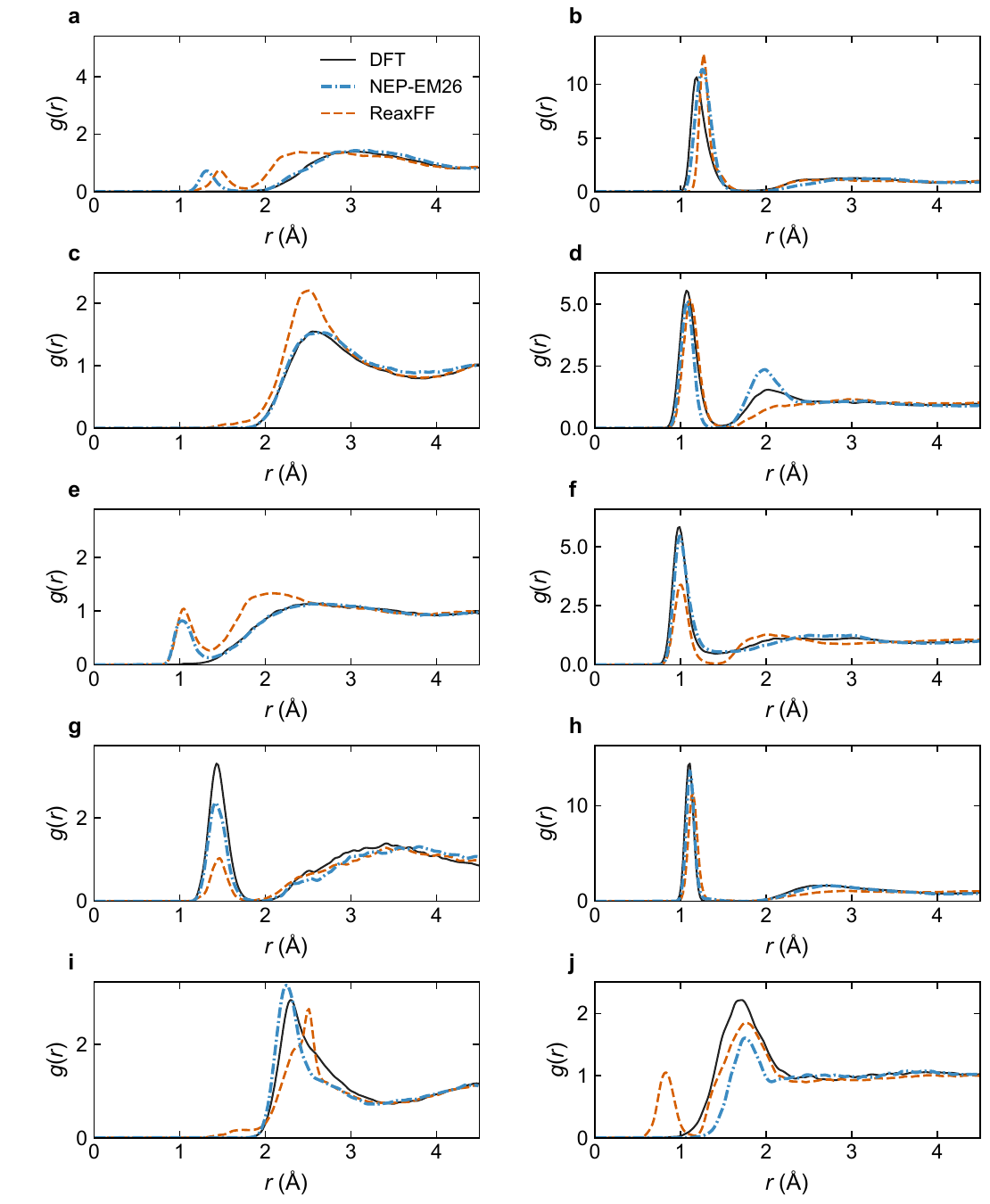}
\caption{All ten C/H/N/O partial RDFs for the selected RDX comparison:
\textbf{a}, C--N; \textbf{b}, C--O; \textbf{c}, N--O;
\textbf{d}, C--H; \textbf{e}, H--N; \textbf{f}, H--O;
\textbf{g}, C--C; \textbf{h}, N--N; \textbf{i}, O--O; \textbf{j}, H--H.
The analysis uses 499 common frames over 15.968--19.952\,ps and the same
normalization and display broadening as the CL-20 comparison.
Pair order and vertical scales match Fig.~\ref{fig:supp_cl20_rdf}.
DFT is black solid, NEP-EM26 blue dash-dotted, and ReaxFF orange dashed.}
\label{fig:supp_rdx_rdf}
\end{figure}

Minimum-image pair distances were accumulated in 0.02\,\AA{} bins from
0 to 4.5\,\AA{}, normalized by the instantaneous volume and finite pair
counts, and averaged equally over the common frames in
Table~\ref{tab:rdf_trajectories}. All analyzed cells were orthorhombic.
The smallest half-cell length was 4.89\,\AA{}, so the 4.5\,\AA{} cutoff
satisfies the minimum-image restriction for every trajectory. Unlike pairs
use $N_A N_B$ available pairs and like pairs use $N_A(N_A-1)/2$, where
$N_A$ and $N_B$ are the respective element counts. These counts also weight
the ten-pair C/H/N/O composite in main Fig.~3a.
Figures~\ref{fig:supp_cl20_rdf} and \ref{fig:supp_rdx_rdf} show all ten
C/H/N/O partial RDFs for CL-20 and RDX in the same order and with matching
vertical scales for each pair. Plotted curves use Gaussian broadening with
standard deviation 0.03\,\AA{}; metrics use unsmoothed curves.

For pair type $p$, $g_p^{\mathrm{model}}(r)$ and $g_p^{\mathrm{DFT}}(r)$ are
the model and DFT RDFs at distance $r$. The normalized integrated absolute
deviation is
\begin{equation}
\mathrm{NIAD}_p=
\frac{\int_0^{4.5\,\text{\AA}}|g_p^{\mathrm{model}}(r)-g_p^{\mathrm{DFT}}(r)|\,\mathrm{d}r}
{\int_0^{4.5\,\text{\AA}}|g_p^{\mathrm{DFT}}(r)|\,\mathrm{d}r}.
\end{equation}
The unweighted mean over the six C/N/O pairs (C--C, C--N, C--O, N--N,
N--O, and O--O) is 0.101 for NEP-EM26 and 0.290 for ReaxFF
in CL-20, and 0.168 and 0.311 in RDX. These six-pair means
are unweighted; the ten-pair composite in main Fig.~3a is weighted by pair
count. For CL-20, main Fig.~3b,c shows the H--O and O--O RDFs; all ten pairs
are given in Fig.~\ref{fig:supp_cl20_rdf}.
Their respective deviations are
0.085 and 0.043 for NEP-EM26, versus 0.315 and 0.242 for ReaxFF.
Figure~\ref{fig:supp_cl20_rdf} retains the complete ten-pair comparison, including
the short-range C--C peak not reproduced by NEP-EM26 and the H--H correlation.
For H--H, the deviations are 0.221 for NEP-EM26 and 0.181 for ReaxFF.

\subsection{Potential-energy evolution}

The energy traces use the same 2{,}494 samples from
0.008 to 19.952\,ps for both CL-20 and RDX, on a common 8\,fs grid.
ABACUS potential energies were converted from Ry to eV using $1\,\mathrm{Ry}\approx13.6057\,\mathrm{eV}$.
The traces use the saved samples directly, without interpolation.

For each method,
\begin{equation}
\Delta E_\mathrm{pot}(t)=
\frac{E_\mathrm{pot}(t)-E_\mathrm{pot}(t_\mathrm{ref})}{N},
\end{equation}
where $E_\mathrm{pot}$ is total-cell potential energy, $t$ is time,
$t_\mathrm{ref}=0.008$\,ps, and $N$ is the atom count. A positive
$\Delta E_\mathrm{pot}$ denotes an increase relative to that saved reference
frame. The displayed curves are 25-point centered means on the 8\,fs
grid (0.192\,ps between the first and last points), truncated at endpoints;
reported means use the unsmoothed samples.

\begin{figure}[!htbp]
\centering
\includegraphics[width=\textwidth]{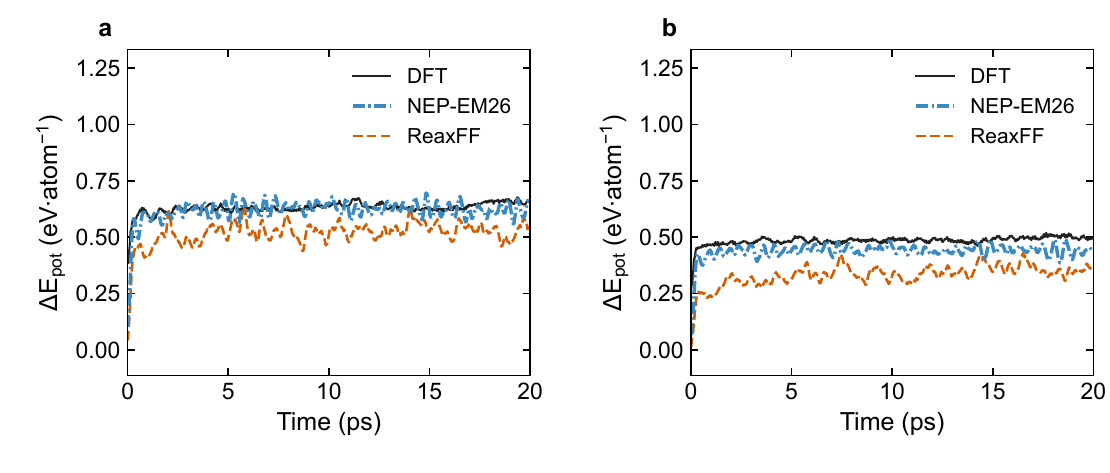}
\caption{Potential-energy changes for \textbf{a}, CL-20, and \textbf{b}, RDX, at model target pressures of 46.27 and 33.37\,GPa, respectively.
Potential-energy change per atom relative to 0.008\,ps, using
2{,}494 common samples over 0.008--19.952\,ps for each material.
Curves show 25-point centered means for visualization only.
DFT is black solid, NEP-EM26 blue dash-dotted, and ReaxFF orange dashed.}
\label{fig:supp_energy_products}
\end{figure}

Figure~\ref{fig:supp_energy_products} compares potential-energy changes
for both materials.
Over the common 15.968--19.952\,ps window, the mean potential-energy changes
for CL-20 are 0.649, 0.619, and 0.527\,eV$\cdot$atom$^{-1}$ for DFT, NEP-EM26,
and ReaxFF.
For RDX they are 0.500, 0.442, and 0.357\,eV$\cdot$atom$^{-1}$.

For the RDF windows in Table~\ref{tab:rdf_trajectories}, the model mean volumes are
1{,}040 and 999\,\AA$^3$ for CL-20 (NEP-EM26 and ReaxFF), versus
1{,}043\,\AA$^3$ for DFT; for RDX they are 1{,}256 and 1{,}184\,\AA$^3$,
versus 1{,}249\,\AA$^3$ for DFT. On these RDF grids, the CL-20 mean pressures
are 46.26, 46.33, and 46.33\,GPa for DFT, NEP-EM26, and ReaxFF;
the RDX means are 33.38, 33.45, and 34.29\,GPa. In particular, the RDX
ReaxFF mean remains about 0.90\,GPa above DFT despite the matched target.
The pressure--volume comparisons are described separately in Section~S3.1.

The final 32.008--40.000\,ps model windows provide a time-window check.
Mean CL-20 volumes are 1{,}039\,\AA$^3$ for NEP-EM26 and
1{,}008\,\AA$^3$ for ReaxFF; RDX values are 1{,}227 and 1{,}190\,\AA$^3$.

\section{D-Hug Calculations and CL-20/RDX State Comparisons}

Pseudo-initial product compositions were obtained by a search that exactly conserves the CHNO stoichiometry of each material. The search maximized the feasible N$_2$ content, minimized isolated atoms when necessary, and then minimized the total molecular reference energy.
The pseudo-initial cells used the atom
counts in Table~\ref{tab:dhug_cells}. For each material, the loading density
$\rho_0$ was used to define the density-rescaled initial volume
$V_\mathrm{load}=V_\mathrm{crystal}\rho_\mathrm{crystal}/\rho_0$, where
$V_\mathrm{crystal}$ is the crystal volume and $\rho_\mathrm{crystal}$ is its
crystal density.
The Packmol~\cite{martinez2009_packmol} cells were generated independently at an atomic number density of
0.10\,atom$\cdot$\AA$^{-3}$.
$E_\mathrm{ref}$, the total reference-state energy (potential plus kinetic), was obtained from a 4\,ps,
300\,K NVT calculation of the unscaled crystal cell using a
Bussi--Donadio--Parrinello thermostat and was
averaged over the final 0.8\,ps. For the eight-material velocity benchmark,
both the reference-state and production calculations used the direct
NEP-EM26 prediction. The model was trained to
PBE-D3(0) labels, so no run-time D3 correction was added. The reference pressure
$P_\mathrm{ref}$ was set to zero.

\begin{sitable}
\caption{Material-specific D-Hug cell sizes and loading densities.}
\label{tab:dhug_cells}
\begin{tabular*}{\sitablewidth}{@{\extracolsep{\fill}}ccc}
\toprule
Material & Atoms & $\rho_0$ (g$\cdot$cm$^{-3}$) \\
\midrule
CL-20   & 144 & 1.980 \\
FOX-7   & 224 & 1.780 \\
HMX    & 224 & 1.890 \\
LLM-105 & 152 & 1.845 \\
PETN   & 116 & 1.763 \\
RDX    & 168 & 1.800 \\
TNT    & 168 & 1.640 \\
Tetryl & 200 & 1.680 \\
\bottomrule
\end{tabular*}
\end{sitable}

Isotropic NPHug trajectories were run in GPUMD at 20, 25, 30, 35, 40, 45,
50, and 55\,GPa. The pressure-coupling period was 2{,}000 steps. Each
trajectory contained 100{,}000 steps with a 0.4\,fs time step, and pressure and
volume were averaged over 32--40\,ps. The eight averaged points were fitted by
unweighted ordinary least squares to
\begin{equation}
P(x)=ax^2+bx+c,
\label{eq:product_hugoniot_quadratic}
\end{equation}
where $P$ is pressure, $x=V/V_\mathrm{load}$ is reduced volume, $V$ is the
cell volume, and $a$, $b$, and $c$ are fitted coefficients with units of
pressure. The Chapman--Jouguet state
was obtained from the tangent Rayleigh line through $(x,P)=(1,0)$ as described
in the main article.

\subsection{CL-20 and RDX pressure--volume comparisons}
\label{sec:supp_dense_pv}

The pressure--volume curves in main Fig.~4 use eight complete
40\,ps trajectories per material and method, with target pressures of
20, 25, 30, 35, 40, 45, 50, and 55\,GPa.
The time step is 0.4\,fs and the output interval 16\,fs.
The final 500 outputs, at 32.016--40.000\,ps, determine each state mean.
For each material, NEP-EM26 and ReaxFF use the same packed initial cell and
atomic coordinates.
NEP-EM26 uses GPUMD NPHug with a pressure-coupling period of
2{,}000 steps; ReaxFF uses an isotropic constant-pressure Hugoniostat integrator within ASE~\cite{larsen2017_ase}, with LAMMPS, the Liu--Goddard
parameters, and stepwise charge equilibration. The ReaxFF Hugoniot constraint
and reported pressure include configurational and kinetic contributions;
the kinetic term is added once to the calculator stress. The reference energy was
obtained separately for each model from a 4\,ps, 300\,K crystal NVT run,
using the final 20\% mean total energy and zero reference pressure.
CL-20 reference crystals were scaled to 2.036\,g$\cdot$cm$^{-3}$, whereas RDX
used the unscaled crystal (1{,}741\,\AA$^3$, 1.695\,g$\cdot$cm$^{-3}$),
which is less dense than the 1.800\,g$\cdot$cm$^{-3}$ loading density. NEP-EM26 reference sampling
used a Bussi--Donadio--Parrinello thermostat~\cite{bussi2007_rescaling}. ReaxFF used fixed-cell LBFGS
preprocessing (stopping at a maximum force of 0.1\,eV$\cdot$\AA$^{-1}$ or
50 steps), followed by Langevin NVT.
The direct NEP-EM26 reference mean total energies were $-936$ and
$-1{,}040$\,eV for the CL-20 and RDX cells, respectively.

The CL-20 reference volume is 1{,}430\,\AA$^3$ at
2.036\,g$\cdot$cm$^{-3}$; the selected DFT reference is 1{,}428\,\AA$^3$
at 2.038\,g$\cdot$cm$^{-3}$. RDX uses 1{,}639\,\AA$^3$ at
1.800\,g$\cdot$cm$^{-3}$ for both. In main Fig.~4, volumes are normalized by
each material's model reference volume $V_0$.
The CL-20 density here differs from the 1.980\,g$\cdot$cm$^{-3}$
used in the eight-material detonation-velocity benchmark.

For each model, the state means were fitted using the same unweighted
quadratic least-squares procedure as in Eq.~\ref{eq:product_hugoniot_quadratic},
with $x=V/V_0$. Curves are shown over each model's sampled volume range.

The ABACUS diagnostic windows use the final 20\% of the available records,
with 12{,}769 outputs for CL-20 and 9{,}980 for RDX.
Their mean pressures and volumes are 46.27\,GPa and
1{,}043\,\AA$^3$ for CL-20, and 33.37\,GPa and
1{,}249\,\AA$^3$ for RDX. These use every 0.4\,fs Hugoniostat
diagnostic record, rather than the 8\,fs post-step coordinate/energy grid
used in Section~S2. DFT error bars show temporal standard deviations of
3.93\,GPa and 18.7\,\AA$^3$ for CL-20, and 2.94\,GPa and
23.5\,\AA$^3$ for RDX, describing fluctuations within each trajectory.

At the displayed DFT volumes, the fitted pressures are 46.38\,GPa for
NEP-EM26 and 41.53\,GPa for ReaxFF in CL-20, and 32.55 and 28.29\,GPa in
RDX.

\subsection{Dispersion-convention sensitivity}
\label{sec:supp_d3_sensitivity}

\begin{figure}[!htbp]
\centering
\includegraphics[width=\textwidth]{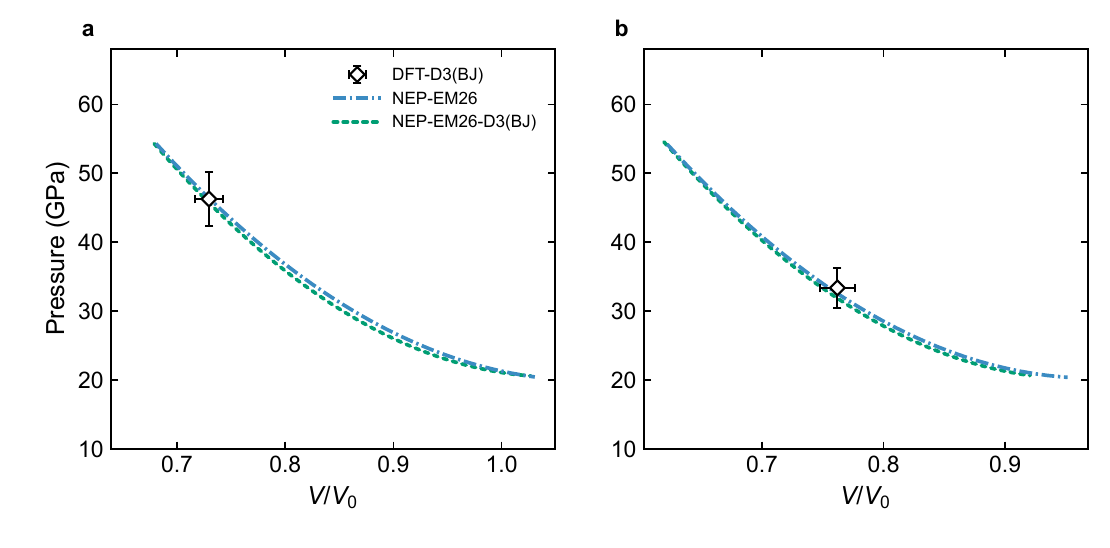}
\caption{Influence of the dispersion convention on the CL-20 and RDX
pressure--volume responses. Results are shown for \textbf{a}, CL-20, and
\textbf{b}, RDX. Curves are quadratic fits to eight trajectory means for
direct NEP-EM26 and NEP-EM26-D3(BJ) within their respective sampled volume
ranges. NEP-EM26-D3(BJ) uses the same learned NEP parameters with the explicit
D3(BJ) minus D3(0) correction applied to the reference and production
calculations. Markers and error bars show the DFT-D3(BJ) trajectory means and
temporal standard deviations; the DFT results are not used in the fits.
Volumes are normalized by a common reference volume $V_0$ for each material.}
\label{fig:supp_d3_sensitivity}
\end{figure}

The influence of the D3(0) and D3(BJ) dispersion conventions on the
NEP-EM26 pressure--volume curves (Fig.~\ref{fig:supp_d3_sensitivity})
was examined using the same eight target
pressures, initial cells, and averaging windows as in main Fig.~4.
NEP-EM26-D3(BJ) was obtained by applying the explicit PBE-D3(BJ) minus
PBE-D3(0) correction~\cite{grimme2010d3,grimme2011_d3bj}:
\begin{equation}
E_\mathrm{pot}^*=E_\mathrm{pot}-E_\mathrm{D3(0)}+E_\mathrm{D3(BJ)},
\end{equation}
where $E_\mathrm{pot}$ is the direct learned total-cell potential energy,
$E_\mathrm{D3(0)}$ and $E_\mathrm{D3(BJ)}$ are the explicit zero-damping
and Becke--Johnson-damping dispersion energies, and $E_\mathrm{pot}^*$ is the
corrected total-cell potential energy. The corresponding force and
configurational-stress differences were applied at every MD step, with
kinetic pressure included once by GPUMD. The corrections used s-dftd3 1.4.0,
the standard PBE parameter sets, pair cutoffs of approximately 94.87 and
95\,bohr for D3(0) and D3(BJ), respectively, and a 40\,bohr
coordination-number cutoff; the three-body Axilrod--Teller--Muto term was
disabled. Both the reference NVT and production trajectories used the
corrected potential. The corrected reference mean total energies were
$-939$ and $-1{,}041$\,eV for the CL-20 and RDX cells, respectively.
The learned NEP parameters were unchanged.

The primary matched-pressure structural and energy comparisons in
Section~\ref{sec:supp_matched_dynamics} and the eight-material
detonation-velocity benchmark used direct NEP-EM26. In separate
matched-pressure trajectories, the correction changed the six-pair RDF
deviation from 0.101 to 0.104 for CL-20 and from 0.168 to 0.247 for RDX.

At the selected DFT states, the fitted volume differences for direct
NEP-EM26 and NEP-EM26-D3(BJ) are $+0.10$\% and $-0.49$\% for CL-20, and
$-0.91$\% and $-1.63$\% for RDX, respectively. For RDX, applying the CJ
construction of Eq.~\ref{eq:product_hugoniot_quadratic} to the corrected
states gives $D=8.595$\,km$\cdot$s$^{-1}$ and
$P_\mathrm{CJ}=35.23$\,GPa, compared with 8.616\,km$\cdot$s$^{-1}$ and
35.63\,GPa for direct NEP-EM26 (Table~\ref{tab:nep_cj_uncertainty}).

\subsection{ReaxFF detonation-velocity estimates}

To match the pressure grid used for the main NEP-EM26 velocity benchmark,
the ReaxFF CJ analysis uses the eight target pressures of 20, 25, 30,
35, 40, 45, 50, and 55\,GPa. All 64 trajectories for the eight reported materials
completed 100{,}000 steps at 0.4\,fs. The final 500 outputs at
32.016--40.000\,ps provide each pressure--volume state.

Each eight-state locus was fitted using the same quadratic least-squares
procedure in Eq.~\ref{eq:product_hugoniot_quadratic}, with
$x=V/V_\mathrm{load}$. The tangent through $(1,0)$ obeys
\begin{equation}
P'(x)=\frac{P(x)}{x-1}, \qquad ax^2-2ax-(b+c)=0.
\end{equation}
Here $P'(x)$ is the derivative of fitted pressure with respect to reduced
volume. A reported solution must be real, satisfy $0<x<1$ and $P(x)>0$,
and lie within that model's sampled $x$ interval. With tangent slope
$m=P(x)/(x-1)<0$ and loading density $\rho_0$, the velocity is
$D=\sqrt{-m/\rho_0}$; GPa and g$\cdot$cm$^{-3}$ give $D$ in km$\cdot$s$^{-1}$.
$P_\mathrm{CJ}$ denotes the pressure at this tangent state.
The same CJ selection criteria were used for NEP-EM26.

ReaxFF used the Liu--Goddard parameters, \texttt{lgvdw yes}, and charge
equilibration every step with a $10^{-6}$ tolerance and 10\,\AA{} cutoff.
Reference energies were obtained independently for each material from the
final 0.8\,ps of 4\,ps, 300\,K Langevin NVT trajectories, following fixed-cell
LBFGS preprocessing limited to 50 steps. The 0.1\,eV$\cdot$\AA$^{-1}$ force
criterion was not reached for any material; the molecules remained intact, and the potential
energy had leveled off by the second half of each NVT run. The reference
pressure was zero. The original crystal cell
was retained for all eight materials, with the loading reference volume
specified separately by the density in main Table~2. In particular, CL-20
used a fresh NVT reference at an unscaled crystal volume of
1{,}430\,\AA$^3$ and a loading reference volume of
1{,}470\,\AA$^3$ at 1.980\,g$\cdot$cm$^{-3}$.
Each force field uses its own reference energy (potential plus kinetic).

Production used the isotropic constant-pressure Hugoniostat integrator within ASE~\cite{larsen2017_ase}, with LAMMPS
force and stress evaluations and with \texttt{tau\_t}=100 and \texttt{tau\_p}=2000 steps.
Both the Hugoniot energy constraint
and reported pressure include configurational and kinetic contributions.
LAMMPS provides configurational stress during force evaluations; ASE adds
the kinetic stress once. Temperature uses $3N$ degrees of freedom, where
$N$ is atom count, and the target-temperature floor is 300\,K.
All eight materials used the same initial product cells and atomic
coordinates as the corresponding NEP-EM26 scans. NEP-EM26
used a Bussi--Donadio--Parrinello reference
thermostat~\cite{bussi2007_rescaling} and GPUMD production integration.
For RDX, the ReaxFF rerun in Section~\ref{sec:supp_dense_pv} uses the same
product cell, loading density, reference energy, and averaging window, with
the initial-velocity setting of the NEP-EM26 scans. Its eight states give
$D=8.123$\,km$\cdot$s$^{-1}$ and $P_\mathrm{CJ}=27.55$\,GPa, compared with
8.149\,km$\cdot$s$^{-1}$ and 28.23\,GPa in Table~\ref{tab:reaxff_cj}.

\begin{sitable}
\caption{ReaxFF CJ analysis of the eight-pressure data for the eight reported materials.
Loading densities match main Table~2. Pressure fit RMSE is calculated
over the eight trajectory means.}
\label{tab:reaxff_cj}
\begin{tabular*}{\sitablewidth}{@{\extracolsep{\fill}}ccccc}
\toprule
Material & $\rho_0$ (g$\cdot$cm$^{-3}$) & $D$ (km$\cdot$s$^{-1}$) & $P_\mathrm{CJ}$ (GPa) & Fit RMSE (GPa) \\
\midrule
CL-20 & 1.980 & 8.487 & 35.08 & 0.432 \\
FOX-7 & 1.780 & 7.832 & 25.33 & 0.595 \\
HMX & 1.890 & 8.593 & 32.14 & 0.773 \\
LLM-105 & 1.845 & 7.726 & 23.79 & 0.322 \\
PETN & 1.763 & 7.358 & 23.51 & 1.137 \\
RDX & 1.800 & 8.149 & 28.23 & 0.795 \\
TNT & 1.640 & 6.580 & 22.11 & 1.876 \\
Tetryl & 1.680 & 6.650 & 22.01 & 0.805 \\
\bottomrule
\end{tabular*}
\end{sitable}

All eight materials in Table~\ref{tab:reaxff_cj} satisfy the sampled physical
tangent criterion at the listed loading densities. These same-density comparisons
have a mean absolute percentage error of 7.4\% for ReaxFF,
versus 2.4\% for NEP-EM26 on the same materials. Errors are computed
from unrounded velocities. LLM-105 favors ReaxFF (3.3\% versus 4.4\%);
the other seven favor NEP-EM26.
Leave-one-pressure-out checks retain a sampled physical tangent in all eight
refits for seven materials. For TNT, seven refits give 6.500--6.893\,km$\cdot$s$^{-1}$,
but omitting the 55\,GPa state removes the sampled physical tangent.
Across these 64 trajectories, the maximum absolute tail-mean Hugoniot residual
is 0.183\,eV$\cdot$atom$^{-1}$, and the maximum volume change between the two halves
of the tail window is 2.65\%, indicating residual finite-time evolution.

\subsection{NEP-EM26 CJ diagnostics and sampling uncertainty}
\label{sec:nep_cj_diagnostics}

The NEP-EM26 diagnostics use the 64 trajectories underlying main Table~2:
eight materials at eight target pressures of 20--55\,GPa. Each state is
averaged over the final 8\,ps, comprising 500 outputs at 0.016\,ps intervals.
Pressure includes configurational and kinetic contributions.
The quadratic fit and physical-tangent criteria are the same as those
described above. Table~\ref{tab:nep_cj_uncertainty} reports the resulting
velocities, CJ pressures, and pressure-fit RMSEs.

Finite-time sampling uncertainty was estimated by a paired circular
moving-block bootstrap. Within each trajectory, pressure and reduced volume
were resampled together in contiguous blocks, with wraparound at the end
of the 500-output window. The primary block length was 64 outputs
(1.024\,ps). Each resample retained 500 outputs, with the final block
truncated as needed. The eight pressure trajectories were resampled
independently; their mean states were then refitted and the CJ tangent
recalculated. For each material, all 10{,}000 resamples yielded a valid
tangent. The 2.5th and 97.5th percentiles define the 95\% sampling intervals
in Table~\ref{tab:nep_cj_uncertainty}. Block lengths of 16, 32, and 125
outputs (0.256, 0.512, and 2.000\,ps) were also tested. Across all four
block lengths and eight materials, the velocity-interval half-widths range
from 0.011 to 0.032\,km$\cdot$s$^{-1}$, with all resampled tangents valid.

\begin{sitable}
\caption{NEP-EM26 CJ estimates and 95\% bootstrap sampling intervals
(brackets) for the eight-pressure benchmark. Loading densities are those
in main Table~2. Pressure-fit RMSE is calculated over the eight trajectory
means; intervals use the primary 1.024\,ps block length.}
\label{tab:nep_cj_uncertainty}
\begin{tabular*}{\sitablewidth}{@{\extracolsep{\fill}}cccc}
\toprule
Material & $D$ (km$\cdot$s$^{-1}$) & $P_\mathrm{CJ}$ (GPa) & Fit RMSE (GPa) \\
\midrule
CL-20 & 9.057 [9.040, 9.074] & 47.63 [46.68, 48.59] & 0.735 \\
FOX-7 & 8.443 [8.426, 8.459] & 34.69 [34.28, 35.13] & 0.827 \\
HMX & 8.942 [8.929, 8.955] & 42.58 [42.07, 43.14] & 0.672 \\
LLM-105 & 8.343 [8.330, 8.355] & 36.64 [36.13, 37.20] & 0.917 \\
PETN & 8.423 [8.399, 8.447] & 33.12 [32.58, 33.71] & 0.858 \\
RDX & 8.616 [8.598, 8.634] & 35.63 [35.17, 36.13] & 0.899 \\
TNT & 7.261 [7.229, 7.292] & 24.78 [24.58, 24.97] & 1.107 \\
Tetryl & 7.476 [7.452, 7.501] & 25.88 [25.67, 26.08] & 0.871 \\
\bottomrule
\end{tabular*}
\end{sitable}

Propagating these resamples to the eight-material mean absolute percentage
error gives 2.4\% with a 95\% sampling interval of 2.35--2.53\%, holding
the literature velocities fixed. As a check on aligned-time dependence,
using shared block indices across all pressure and material traces gives
2.34--2.54\%. These intervals quantify finite-time sampling for the fixed
potential, reference states, densities, and quadratic fit, assuming
approximately stationary tails. Model bias, reference-energy uncertainty,
finite-size effects, independent-seed variation, and uncertainty in the
literature values are outside this estimate.

Sensitivity to the averaging window and pressure grid is reported
separately in Table~\ref{tab:nep_cj_sensitivity}. Using the final 4, 8, or
12\,ps yields valid tangents for all eight materials and changes the velocity
by at most 0.43\% relative to the final-8-ps result. Leave-one-pressure-out
refits give a maximum change of 1.84\% among valid solutions. For CL-20,
omitting the 55\,GPa state places the tangent outside the remaining sampled
volume range; the other seven omissions retain valid tangents. All eight
omissions retain valid tangents for each of the other materials.

\begin{sitable}
\caption{NEP-EM26 CJ sensitivity to averaging-window length and pressure-point
omission. Changes are the maximum absolute percentage changes relative to
the eight-state, final-8-ps velocity. The window column compares the final
4, 8, and 12\,ps; the omission column includes only valid seven-state refits.}
\label{tab:nep_cj_sensitivity}
\begin{tabular*}{\sitablewidth}{@{\extracolsep{\fill}}cccc}
\toprule
Material & \shortstack{Window change\\(\%)} &
\shortstack{Omission change\\(\%)} & \shortstack{Valid omission refits\\(out of eight)} \\
\midrule
CL-20 & 0.08 & 0.26 & 7 \\
FOX-7 & 0.18 & 0.64 & 8 \\
HMX & 0.05 & 0.55 & 8 \\
LLM-105 & 0.15 & 0.60 & 8 \\
PETN & 0.17 & 0.63 & 8 \\
RDX & 0.37 & 0.90 & 8 \\
TNT & 0.43 & 1.84 & 8 \\
Tetryl & 0.30 & 0.99 & 8 \\
\bottomrule
\end{tabular*}
\end{sitable}

The instantaneous Hugoniot residual per atom was evaluated as
\begin{equation}
r_\mathrm{H}(t)=\frac{1}{N}\left\{
E(t)-E_\mathrm{ref}
-\frac{P(t)+P_\mathrm{ref}}{2}\left[V_\mathrm{load}-V(t)\right]
\right\},
\label{eq:nep_cj_residual}
\end{equation}
where $E$ is total cell energy, $N$ is atom count, and pressure--volume
products are converted to energy units. Table~\ref{tab:nep_cj_drift} reports
the maximum absolute tail-mean residual and half-window changes for each
material. The latter compare the two 4\,ps halves of the final 8\,ps:
$\Delta P=\overline{P}_2-\overline{P}_1$,
$\Delta V=100\%\cdot(\overline{V}_2/\overline{V}_1-1)$, and
$\Delta T=100\%\cdot(\overline{T}_2/\overline{T}_1-1)$.
Across all 64 trajectories, the largest absolute mean residual is
8.93\,meV$\cdot$atom$^{-1}$ and the largest volume change is 1.46\%.

\begin{sitable}
\caption{NEP-EM26 Hugoniot residuals and half-window changes. Each entry is
the maximum absolute value over the eight pressure trajectories for that
material; column maxima can occur at different pressures.
$\langle r_\mathrm{H}\rangle$ is averaged over the final 8\,ps.}
\label{tab:nep_cj_drift}
\begin{tabular*}{\sitablewidth}{@{\extracolsep{\fill}}ccccc}
\toprule
Material & \shortstack{$|\langle r_\mathrm{H}\rangle|$\\(meV$\cdot$atom$^{-1}$)} &
$|\Delta P|$ (GPa) & $|\Delta V|$ (\%) & $|\Delta T|$ (\%) \\
\midrule
CL-20 & 5.82 & 0.36 & 1.39 & 2.89 \\
FOX-7 & 2.46 & 0.19 & 0.63 & 3.69 \\
HMX & 4.44 & 0.22 & 0.87 & 2.74 \\
LLM-105 & 4.69 & 0.35 & 0.81 & 5.26 \\
PETN & 8.93 & 0.39 & 1.00 & 3.15 \\
RDX & 5.74 & 0.19 & 1.46 & 3.22 \\
TNT & 4.84 & 0.66 & 0.86 & 3.59 \\
Tetryl & 2.98 & 0.37 & 0.87 & 3.11 \\
\bottomrule
\end{tabular*}
\end{sitable}

\section{Computational-Efficiency Benchmark}
\label{sec:supp_efficiency}

The benchmark compares NEP-EM26 in GPUMD 5.5~\cite{fan2022gpumd,gpumd4}
with the KOKKOS-accelerated ReaxFF implementation in LAMMPS~\cite{thompson2022_lammps}
(30 March 2026, revision \texttt{f2f2286}), using the
\texttt{ffield.reax.lg} parameterization~\cite{liu2011_reaxfflg}.
Both methods used the same NVIDIA RTX PRO 6000 Blackwell Server Edition
GPU, with no concurrent GPU workloads, one CPU core, and one host thread.
ReaxFF used one MPI process. Runs used a
microcanonical (NVE) ensemble, velocities initialized at 300\,K,
a 0.25\,fs time step, and a 20-step warm-up.
The NEP-EM26 parameters were the same as in the accuracy benchmarks
(Table~\ref{tab:nep_settings}).
NEP-EM26 timings used the direct GPUMD implementation of the learned potential,
with no additional run-time dispersion term.
ReaxFF-lg used a 10\,\AA{} nonbonded cutoff and the default
\texttt{thb\_cutoff} of 0.001. Charge equilibration~\cite{rappe1991_qeq}
was performed every step with \texttt{qeq/reaxff/kk}, a 0--10\,\AA{}
taper range, a tolerance of $10^{-6}$, and a maximum of 200 iterations.
The neighbor-list skin was 2\,\AA{}, with a rebuild check every step.

The timing structures were constructed from a 20-atom molecular-crystal cell
containing two C$_2$H$_4$N$_2$O$_2$ molecules at a density of
1.565\,g$\cdot$cm$^{-3}$. The 160-atom structure used an eightfold supercell
with lattice vectors $2\mathbf{a}+2\mathbf{b}$,
$-2\mathbf{a}+2\mathbf{b}$, and $\mathbf{c}$, where $\mathbf{a}$,
$\mathbf{b}$, and $\mathbf{c}$ are the reference-cell vectors.
This choice gives perpendicular cell heights greater than the 10\,\AA{} cutoff.
The remaining structures were generated by $n\times n\times n$ replication
with $n=3,4,5,6,7,10,12,15,17,18,19,20$.
Both methods used the same structure at each size, covering thirteen system
sizes from 160 to 160{,}000 atoms.

For each method and size, a calibration run
set the production length to target approximately 10\,s of runtime, within
bounds of 20--5{,}000 steps. Ten separate process runs were then performed
per method and size, alternating the methods. The time per step was the production-loop wall time divided by its
step count, excluding initialization and warm-up. All 260 runs completed
without nonfinite output or QEq convergence warnings and passed the
5\,meV$\cdot$atom$^{-1}$ sampled NVE energy-change screen.
Table~\ref{tab:timing_results} reports the production lengths and mean
step times. The 95\% confidence intervals (CIs) were calculated using Student's
$t$ distribution as $\bar{t}\pm t_{0.975,9}\,s/\sqrt{10}$, where $s$ is the sample standard
deviation across the ten runs; they describe repeat-to-repeat timing
variability.

Across the thirteen sizes, NEP-EM26 mean step times range from
1.11 to 6.01\,ms$\cdot$step$^{-1}$, compared with
25.1--156.6\,ms$\cdot$step$^{-1}$ for ReaxFF
(main Fig.~5). The ReaxFF-to-NEP-EM26 ratios of mean
step times are 15.8--26.0, exceeding one order of magnitude at every tested size.

\begin{sitable}
\caption{Mean time per molecular dynamics step and 95\% confidence-interval
half-widths from ten runs per method and size. Times are in
ms$\cdot$step$^{-1}$. Production steps exclude the 20-step warm-up.
The ratio is ReaxFF mean time divided by NEP-EM26 mean time.}
\label{tab:timing_results}
\fontsize{8}{10}\selectfont
\setlength{\tabcolsep}{3pt}
\begin{tabular*}{\sitablewidth}{@{\extracolsep{\fill}}rrrrrr}
\toprule
 & \multicolumn{2}{c}{Production steps} & \multicolumn{2}{c}{Mean time $\pm$ 95\% CI half-width} & \\
\cmidrule(lr){2-3}\cmidrule(lr){4-5}
Atoms & NEP-EM26 & ReaxFF & NEP-EM26 & ReaxFF & Ratio \\
\midrule
160 & 5{,}000 & 360 & $1.106 \pm 0.005$ & $25.09 \pm 0.12$ & 22.7 \\
540 & 5{,}000 & 360 & $1.416 \pm 0.005$ & $25.53 \pm 0.23$ & 18.0 \\
1{,}280 & 5{,}000 & 380 & $1.613 \pm 0.009$ & $26.06 \pm 0.11$ & 16.2 \\
2{,}500 & 5{,}000 & 360 & $1.659 \pm 0.004$ & $26.50 \pm 0.08$ & 16.0 \\
4{,}320 & 5{,}000 & 380 & $1.674 \pm 0.009$ & $26.42 \pm 0.10$ & 15.8 \\
6{,}860 & 5{,}000 & 320 & $1.686 \pm 0.006$ & $30.04 \pm 0.27$ & 17.8 \\
20{,}000 & 5{,}000 & 280 & $1.947 \pm 0.006$ & $34.55 \pm 0.14$ & 17.7 \\
34{,}560 & 4{,}520 & 200 & $2.279 \pm 0.009$ & $50.68 \pm 0.22$ & 22.2 \\
67{,}500 & 2{,}980 & 120 & $3.226 \pm 0.018$ & $75.91 \pm 0.27$ & 23.5 \\
98{,}260 & 1{,}820 & 80 & $4.988 \pm 0.031$ & $110.14 \pm 0.49$ & 22.1 \\
116{,}640 & 1{,}820 & 80 & $5.358 \pm 0.022$ & $121.81 \pm 0.24$ & 22.7 \\
137{,}180 & 1{,}580 & 80 & $5.657 \pm 0.028$ & $131.57 \pm 0.59$ & 23.3 \\
160{,}000 & 1{,}560 & 60 & $6.013 \pm 0.056$ & $156.56 \pm 0.66$ & 26.0 \\
\bottomrule
\end{tabular*}
\end{sitable}

\section{Model, Dataset, and DFT Details}

\subsection{NEP-EM26 input and optimization}

Table~\ref{tab:nep_settings} lists the model and optimization settings used for
the reported potential. NEP-EM26 was trained on the PBE-D3(0) reference labels
with the separable natural evolution
strategy (SNES) in two consecutive stages with the same architecture. The first stage started from randomly initialized parameters and ran for 200{,}000 generations with a batch size of 2{,}000 configurations. The second stage continued from the final first-stage parameters and their per-parameter SNES search widths and ran for another 200{,}000 generations, using the full training set as a single batch. The model snapshot with the best performance on the
validation set was selected for all reported NEP-EM26 predictions.

\begin{sitable}
\caption{NEP-EM26 architecture and optimization settings. Superscripts
$\mathrm{R}$ and $\mathrm{A}$ denote radial and angular descriptor channels,
respectively; $n_\mathrm{max}^{\mathrm{R}}$ and $n_\mathrm{max}^{\mathrm{A}}$ are
the maximum radial expansion orders, $N_\mathrm{bas}^{\mathrm{R}}$ and
$N_\mathrm{bas}^{\mathrm{A}}$ are the corresponding upper basis indices
(the GPUMD \texttt{basis\_size} parameters); a value of 8 includes nine
basis functions indexed from 0 to 8,
and $l_\mathrm{max}^{3\mathrm{b}}$, $l_\mathrm{max}^{4\mathrm{b}}$, and
$l_\mathrm{max}^{5\mathrm{b}}$ are the maximum angular orders for the three-,
four-, and five-body channels, respectively. The $\lambda$ values are the
relative energy, force, and stress-related loss weights. Values separated by a slash refer to the first and second training stages, respectively.}
\label{tab:nep_settings}
\begin{tabular*}{\sitablewidth}{@{\extracolsep{\fill}}cc}
\toprule
Setting & Value \\
\midrule
Elements & C, H, N, O \\
Radial cutoff & 5.0\,\AA{} \\
Angular cutoff & 4.0\,\AA{} \\
$n_\mathrm{max}^{\mathrm{R}},n_\mathrm{max}^{\mathrm{A}}$ & 6, 6 \\
$N_\mathrm{bas}^{\mathrm{R}},N_\mathrm{bas}^{\mathrm{A}}$ & 8, 8 \\
$l_\mathrm{max}^{3\mathrm{b}},l_\mathrm{max}^{4\mathrm{b}},l_\mathrm{max}^{5\mathrm{b}}$ & 4, 2, 0 \\
Hidden layers & One per element type \\
Neurons per hidden layer & 70 \\
Activation & Hyperbolic tangent \\
ZBL inner and outer radii & 1 and 2\,\AA{} \\
Optimizer & Separable natural evolution strategy \\
Batch size (stage 1 / stage 2) & 2{,}000 / 65{,}209 (full training set) \\
Generations (stage 1 / stage 2) & 200{,}000 / 200{,}000 \\
Population size & 50 \\
Total optimized parameters & 14{,}337 \\
L1 and L2 regularization weights (each) & 0.0599 \\
Model selection & Best validation-set performance \\
$\lambda_E:\lambda_F:\lambda_V$ (stage 1 / stage 2) & $1.0:1.0:0.1$ / $5.0:1.0:1.0$ \\
\bottomrule
\end{tabular*}
\end{sitable}

For the stress-related fitting term, GPUMD uses the normalized tensor
$\boldsymbol{\sigma}V/N$, where $\boldsymbol{\sigma}$ is compression-positive
configurational stress in eV$\cdot$\AA$^{-3}$, $V$ is the cell volume in \AA$^3$, and $N$ is the
atom count. The held-out stress errors in Section~S1 are evaluated after
conversion to GPa.

\subsection{Dataset collection and configuration selection}

\subsubsection{CSD structures}

Crystallographic Information Files containing only C, H, N, and O were
converted to VASP format. Structures containing more than 500 atoms were
discarded, and a minimum-image distance filter retained structures with a
minimum interatomic distance between 0.5 and 2.5\,\AA. Structures containing
elements outside CHNO were removed. Near duplicates were identified with
pymatgen~\cite{ong2013_pymatgen} \texttt{StructureMatcher} using lattice, site, and angle tolerances of
0.15, 0.25, and $5^\circ$, respectively; comparisons were grouped by atom count
and nitrogen mass fraction. Primitive-cell reduction and volume rescaling were
enabled, whereas supercell matching and subset matching were disabled. A final
density filter retained structures with density greater than
0.8\,g$\cdot$cm$^{-3}$. The 55{,}171 labeled CSD candidates were then screened using
their reference labels. Configurations were retained only when the energy per atom was
no greater than $-4.0$\,eV$\cdot$atom$^{-1}$, the maximum atomic force norm was no
greater than 120\,eV$\cdot$\AA$^{-1}$, and every compression-positive stress component lay between
$-0.4$ and $1.1$\,eV$\cdot$\AA$^{-3}$. These criteria had to be satisfied by both
the D3-corrected and uncorrected reference labels for the same configuration.
Of the 55{,}171 labeled CSD candidates, 54{,}574 passed this screen and were
retained in the final dataset (main Fig.~6a).

\subsubsection{Reactive-MD farthest-point sampling}

The final source-resolved candidate pool contains 182{,}122
configurations distributed over eleven material systems and twelve trajectory
branches. For configuration $i$, the preliminary NEP supplied an atomic
descriptor matrix $\mathbf{Q}_i$, whose $j$th row, denoted $\mathbf{q}^{(i)}_j$, is the
descriptor vector for atom $j$. Let $N_i$ be the number of atoms in
configuration $i$. The structure-level vector used for sampling was defined
as follows:
\begin{equation}
\bar{\mathbf{q}}_i=\frac{1}{N_i}\sum_{j=1}^{N_i}\mathbf{q}^{(i)}_j,
\end{equation}
where $\bar{\mathbf{q}}_i$ is the descriptor vector for configuration $i$.
The \texttt{FarthestPointSample} selector from PyNEP was then applied directly
to these vectors using Euclidean distance. With no preselected seed set, the
implementation starts from the first input vector, repeatedly adds the
configuration whose distance to its nearest selected vector is greatest, and
stops when this maximum nearest-neighbor distance no longer exceeds the chosen
threshold.

The descriptive principal component projection in main Fig.~6b uses all
81{,}512 retained configurations. Its eight input features are mass density,
pressure, energy per atom, volume per atom, and the C/H/N/O atomic fractions.
Energy and pressure were extracted from the final reference data used for
training, validation, and testing. Pressure was obtained as the mean of the
three diagonal compression-positive stress components. Each feature was centered and
scaled to unit population standard deviation over the complete dataset before
principal component analysis (PCA). The first two components explain 56.9\% and
19.8\% of the variance, respectively. This PCA visualizes physical and
compositional features; farthest-point sampling uses the NEP descriptors
defined above.

During method development, a global threshold scan was performed on an
assembled 185{,}736-configuration pool; Table~\ref{tab:fps_thresholds} reports
the numbers retained in that sensitivity test. For final configuration selection, the
same minimum distance of 0.007 was applied independently within each source
branch of the 182{,}122-configuration pool, retaining 27{,}008 configurations
in total. Of the 27{,}008 single-point DFT jobs, 26{,}938 produced complete
energy, force, and stress records and form the reactive-MD component
(Table~\ref{tab:reactive_sources}). The
remaining 70 jobs terminated during VASP initialization, before the first
electronic self-consistent-field iteration, and therefore returned no final
energy. They were excluded from the final reactive-MD component.

\begin{sitable}
\caption{Global farthest-point sampling (FPS) threshold sensitivity for the 185{,}736-configuration
pool; the final source-wise pass used 0.007.}
\label{tab:fps_thresholds}
\begin{tabular*}{\sitablewidth}{@{\extracolsep{\fill}}cc}
\toprule
Minimum distance & Retained configurations \\
\midrule
0.0500 & 94 \\
0.0250 & 338 \\
0.0100 & 5{,}637 \\
0.0085 & 11{,}717 \\
0.0075 & 20{,}295 \\
0.0070 & 27{,}168 \\
0.0065 & 36{,}421 \\
0.0050 & 79{,}717 \\
\bottomrule
\end{tabular*}
\end{sitable}

\begin{sitable}
\caption{Source-resolved composition of the final reactive-MD component.}
\label{tab:reactive_sources}
\begin{tabular*}{\sitablewidth}{@{\extracolsep{\fill}}cccc}
\toprule
Source & Configurations & Source & Configurations \\
\midrule
cgN    & 3{,}862 & CL-20  & 3{,}384 \\
HNS    & 3{,}124 & 5AT   & 3{,}081 \\
LLM-105 & 3{,}059 & DNAN  & 2{,}257 \\
1HT    & 2{,}186 & TKX50 & 1{,}901 \\
HMX    & 1{,}579 & NTO   & 1{,}512 \\
FOX-7   &   993 & Total & 26{,}938 \\
\bottomrule
\end{tabular*}
\end{sitable}

The reactive-MD pressure distribution in main Fig.~6d uses all 26{,}938
retained configurations. For the temperature distribution in main Fig.~6e,
source thermodynamic records were retained only when the absolute pressure
difference between the configuration metadata and the matched record was
at most 0.01\,GPa. The temperature distribution therefore contains 26{,}850
configurations; the 88 unmatched configurations remain in the final dataset
and its split. The held-out error map uses the separate matching criteria
in Section~S1.2.

The final 81{,}512 configurations were randomly divided at the configuration
level into 65{,}209 training, 8{,}152 validation, and 8{,}151 test
configurations. The split was generated by shuffling the configuration indices
with NumPy's \texttt{default\_rng} using seed 20260606. The first 65{,}209
indices were assigned to training, the next 8{,}151 to testing, and the
remaining 8{,}152 to validation. Farthest-point sampling, configuration and label
curation, and CSD label-quality screening were completed before this split. The
8{,}151-configuration test partition was used neither for parameter updates
nor for model selection, which was based on the validation set; its errors
were only logged during optimization.

\subsection{DFT reference labels and dispersion convention}

Single-point (\texttt{NSW}=0) PBE-D3(0) labels were generated with VASP 6.3.0~\cite{kresse1996vasp} and
PAW-PBE~\cite{kresse1999_paw,perdew1996pbe} datasets for C, H, N, and O. The settings were a 700\,eV cutoff, a
$\Gamma$-centered grid with \texttt{KSPACING}=0.7\,\AA$^{-1}$,
\texttt{PREC=Accurate}, \texttt{ALGO=Fast}, an electronic threshold of
$10^{-6}$\,eV, at most 100 electronic steps, \texttt{ISMEAR}=0,
$\sigma=0.05$\,eV, \texttt{LREAL=False}, \texttt{LASPH=False}, and zero-damping
D3(0)~\cite{grimme2010d3} (\texttt{IVDW}=11). The resulting energy, force, and stress labels were used
for training and held-out parity. Except for calculations explicitly labeled
NEP-EM26-D3(BJ) in Section~\ref{sec:supp_d3_sensitivity}, the reported NEP-EM26 calculations used the
direct learned potential without an additional run-time D3 term.

\FloatBarrier
% Shared database; only references cited in this SI are included.
\begingroup
\urlstyle{same}
\apptocmd{\thebibliography}{\interlinepenalty=10000}{}{}
\bibliography{references}
\endgroup